\documentclass[twocolumn,showpacs,preprintnumbers,amsmath,amssymb]{revtex4}

\usepackage{graphicx}
\usepackage{dcolumn}
\usepackage{bm}

\newcommand{\bq}{\begin{equation}}
\newcommand{\eq}{\end{equation}}
\newcommand{\bqa}{\begin{eqnarray}}
\newcommand{\eqa}{\end{eqnarray}}
\newcommand{\nn}{\nonumber \\}
\newcommand{\ij}{\langle i j \rangle}

\def\be     {\begin{equation}}
\def\ee     {\end{equation}}
\def\bea        {\begin{eqnarray}}
\def\eea        {\end{eqnarray}}
\def\bnn    {\begin{eqnarray*}}
\def\enn    {\end{eqnarray*}}

\begin{document}

\title{How to control pairing fluctuations: SU(2) slave-rotor
gauge theory of the Hubbard model }
\author{Ki-Seok Kim}
\affiliation{ School of Physics, Korea Institute for Advanced
Study, Seoul 130-012, Korea }
\date{\today}

\begin{abstract}
We study how to incorporate Mott physics in the BCS-type
superconductor, motivated from the fact that high $T_c$
superconductivity results from a Mott insulator via hole doping.
The U(1) slave-rotor representation was proposed to take local
density fluctuations into account non-perturbatively, describing
the Mott-Hubbard transition at half filling. Since this
decomposition cannot control local pairing fluctuations, the U(1)
slave-rotor representation does not give a satisfactory treatment
for charge fluctuations. Extending the U(1) slave-rotor
representation, we introduce an SU(2) slave-rotor representation
to allow not only local density fluctuations but also local
pairing excitations. We find an SU(2) slave-rotor gauge theory of
the Hubbard model in terms of two kinds of collective boson
excitations associated with density and pairing fluctuations that
interact with gapless fermion excitations via SU(2) gauge
fluctuations. An interesting observation in this effective
description is that phase fluctuations of fermion pairs arise as
SU(2) gauge fluctuations. Thus, fermion-pairing excitations can be
controlled by dynamics of collective bosons in the SU(2)
slave-rotor gauge theory. Performing the standard saddle-point
analysis based on the SU(2) slave-rotor action, we find an
interesting phase described by partial freezing of charge
fluctuations near the Mott-Hubbard critical point, where local
density-fluctuation modes are condensed but local pair-excitation
modes are gapped. Partial freezing of charge fluctuations causes
fermion pairing to be incoherent as a result of reconciliation of
superconductivity and Mott physics. The nature of this
non-superconducting phase is identified with an anomalous metal
due to the presence of incoherent pairing.
\end{abstract}


\maketitle

\section{Introduction}

Nature of the doped Mott insulator is one of the central interests
in modern condensed matter physics, associated with the mechanism
of high $T_c$ superconductivity. Considering the high $T_c$ phase
diagram, an antiferromagnetic order in the parent Mott insulator
vanishes rapidly via hole doping, and a paramagnetic
non-superconducting phase appears. This non-magnetic state evolves
into the high $T_c$ superconductor, doping holes
further.\cite{RMP} The central issue is the nature of the
intermediate non-magnetic phase between the antiferromagnetic Mott
insulator and high $T_c$ superconductor. It is important to notice
that physics of the superconducting state is BCS-like although the
high $T_c$ superconductivity results from the doped Mott
insulator.\cite{RMP} In this respect the intermediate phase will
be determined by competition of Mott physics and BCS
superconductivity. In this paper we discuss how to incorporate the
Mott physics into the BCS superconductivity.

To understand the Mott physics in a concrete manner, we consider
the BCS-Hubbard Hamiltonian \bqa && H = -
t\sum_{\ij\sigma}(c_{i\sigma}^{\dagger}c_{j\sigma} + H. c.) \nn &&
-
\sum_{\ij}|\Delta_{ij}|e^{-ia_{ij}}(c_{i\uparrow}^{\dagger}c_{j\downarrow}^{\dagger}
- c_{i\downarrow}^{\dagger}c_{j\uparrow}^{\dagger}) - H. c. +
\frac{1}{J}\sum_{\ij}|\Delta_{ij}|^{2} \nn && +
U\sum_{i}n_{i\uparrow}n_{i\downarrow} , \eqa where $\Delta_{ij}$
is the pairing potential with its amplitude $|\Delta_{ij}|$ and
phase $a_{ij}$. This Hamiltonian introduces the competing nature
arising from the density-phase uncertainty; the $J$ term causes
local pairing of electrons while the Hubbard-$U$ term suppresses
local charge fluctuations, thus breaking phase coherence of
electron pairs. When local charge fluctuations are strong in the
case of $U < U_{c}$ with the critical strength $U_{c}$ for the
Mott transition, phase fluctuations of electron pairs would be
suppressed, i.e., $\langle{e}^{-ia_{ij}}\rangle = 1$ owing to the
density-phase uncertainty relation. In this case superconductivity
is expected to appear, described by the BCS model $L_{SC} =
\sum_{i\sigma}c_{i\sigma}^{\dagger}(\partial_{\tau} - \mu
)c_{i\sigma} - t\sum_{\ij\sigma}(c_{i\sigma}^{\dagger}c_{j\sigma}
+ H. c.) -
\sum_{\ij}|\Delta_{ij}|(c_{i\uparrow}^{\dagger}c_{j\downarrow}^{\dagger}
- c_{i\downarrow}^{\dagger}c_{j\uparrow}^{\dagger}) - H. c.$ On
the other hand, when local density fluctuations are suppressed in
the case of $U > U_{c}$, the density-phase uncertainty causes
$\langle{e}^{-ia_{ij}}\rangle = 0$. The resulting
non-superconducting phase would be described by the effective
Lagrangian $L_{eff} =
\sum_{i\sigma}c_{i\sigma}^{\dagger}(\partial_{\tau} - \mu
)c_{i\sigma} - t\sum_{\ij\sigma}(c_{i\sigma}^{\dagger}c_{j\sigma}
+ H. c.) -
\sum_{\ij}|\Delta_{ij}|{e}^{-ia_{ij}}(c_{i\uparrow}^{\dagger}c_{j\downarrow}^{\dagger}
- c_{i\downarrow}^{\dagger}c_{j\uparrow}^{\dagger}) - H. c. -
\frac{1}{g^2}\sum_{\Box}\cos(\partial\times{a})$, where the gauge
action is introduced to impose $\langle{e}^{-ia_{ij}}\rangle = 0$.
This state can be regarded as a non-Fermi liquid metal with
phase-incoherent pairs.\cite{Trivedi,Laughlin}

The problem is how to control local charge fluctuations. Recently,
U(1) slave-rotor representation was proposed in order to take
local charge fluctuations into account
non-perturbatively.\cite{Florens} Actually, the U(1) slave-rotor
gauge theory of the Hubbard Hamiltonian found the
bandwidth-control Mott transition from spin liquid to Fermi liquid
at half filling, ignoring antiferromagnetism and
superconductivity.\cite{Florens,LeeLee,Kim_SR_disorder} Using the
U(1) slave-rotor decomposition $c_{i\sigma} =
e^{-i\theta_{i}}f_{i\sigma}$, we obtain the following expression
from Eq. (1) \bqa && Z = \int{Df_{i\sigma}D\theta_{i}
D|\Delta_{ij}|Da_{ij}DL_{i}D\varphi_{i}} e^{-\int_{0}^{\beta}d\tau
L} , \nn && L =
\sum_{i\sigma}f_{i\sigma}^{\dagger}(\partial_{\tau} -
\mu)f_{i\sigma} -
t\sum_{\ij\sigma}(f_{i\sigma}^{\dagger}e^{i\theta_{i}}e^{-i\theta_{j}}f_{j\sigma}
+ H. c.) \nn && + \sum_{i}\Bigl( \frac{U}{4}L_{i}^{2} -
iL_{i}\partial_{\tau}\theta_{i} + i\varphi_{i}(L_{i} -
[\sum_{\sigma}{f}^{\dagger}_{i\sigma}{f}_{i\sigma} - 1]) \Bigr)
\nn && -
\sum_{\ij}|\Delta_{ij}|e^{i\theta_{i}}e^{-ia_{ij}}e^{i\theta_{j}}(f_{i\uparrow}^{\dagger}f_{j\downarrow}^{\dagger}
- f_{i\downarrow}^{\dagger}f_{j\uparrow}^{\dagger}) - H. c. \nn &&
+ \frac{1}{J}\sum_{\ij}|\Delta_{ij}|^{2} , \eqa where the
Hubbard-$U$ term is represented as the charge and spin channels
$U\sum_{i}n_{i\uparrow}n_{i\downarrow} =
\frac{U}{4}\sum_{i}(n_{i\uparrow} + n_{i\downarrow} - 1)^{2} -
\frac{U}{4}\sum_{i}(n_{i\uparrow} - n_{i\downarrow})^{2} +
\frac{U}{2}\sum_{i}(n_{i\uparrow} + n_{i\downarrow}) -
\frac{U}{4}\sum_{i}1$, and the spin channel is not taken into
account in this paper. It is easy to show that Eq. (2) is exactly
the same as Eq. (1) with the charge channel for the Hubbard-$U$
term, integrating out the $\varphi_{i}$ and $L_i$ fields with
$f_{i\sigma} = e^{i\theta_{i}}c_{i\sigma}$. In this expression the
electron Hilbert space $|c_{i\sigma}>$ is reconstructed as the
direct product of the boson and fermion Hilbert spaces
${|L_{i}>}\bigotimes{|f_{i\sigma}>}$ according to the
decomposition $c_{i\sigma} = e^{-i\theta_{i}}f_{i\sigma}$, where
$L_{i}$ represents an electron density at site $i$. It is clear
that any decomposition method enlarges the original electron
Hilbert space, thus an appropriate constraint associated with the
decomposition should be imposed. The Lagrange multiplier field
$\varphi_{i}$ expresses the U(1) slave-rotor constraint $L_{i} =
\sum_{\sigma}f_{i\sigma}^{\dagger}f_{i\sigma} - 1$, implying that
the fermion and boson Hilbert spaces are not independent, thus the
two operators $f_{i\sigma}$ and $e^{i\theta_{i}}$ also. Then,
$e^{-i\theta_{i}}$ is identified with an annihilation operator of
an electron charge owing to the constraint $L_{i} =
\sum_{\sigma}f_{i\sigma}^{\dagger}f_{i\sigma} - 1$ and the
canonical relation $[L_{i}, \theta_{j}] = -i\delta_{ij}$ imposed
by $- iL_{i}\partial_{\tau}\theta_{i}$.

Integrating out the density field $L_{i}$ and performing the
Hubbard-Stratonovich (HS) transformation to decompose the
"correlated" hopping term in the following way \bqa && \exp\Bigl[
- \int_{0}^{\beta}d\tau \Bigl\{ -
t\sum_{\ij\sigma}(f_{i\sigma}^{\dagger}e^{i\theta_{i}}e^{-i\theta_{j}}f_{j\sigma}
+ H. c.) \Bigr\}\Bigr] \nn && = \int{D\alpha_{ij}D\beta_{ij}}
\exp\Bigl[ - \int_{0}^{\beta}d\tau \Bigl\{
t\sum_{\ij}(\alpha_{ij}^{*}\beta_{ij} + H.c.) \nn && -
t\sum_{\ij\sigma}(f_{i\sigma}^{\dagger}\alpha_{ij}^{*}f_{j\sigma}
+ H. c.) - t\sum_{\ij}(e^{i\theta_{i}}\beta_{ij}e^{-i\theta_{j}} +
H. c.) \Bigr\} \Bigr] , \nonumber \eqa we obtain the U(1)
slave-rotor gauge Lagrangian with $d-wave$ pairing \bqa && Z =
\int{Df_{i\sigma}D\theta_{i} Da_{ij}D\varphi_{i}
Dc_{ij}}e^{-\int_{0}^{\beta}{d\tau} L} , \nn && L = L_{0} + L_{f}
+ L_{\theta} , ~~~~~~~ L_{0} = \sum_{\ij}(t\alpha\beta +
\frac{\Delta^{2}}{J} ) , \nn && L_{f} =
\sum_{i\sigma}f_{i\sigma}^{\dagger}(\partial_{\tau} -
\mu)f_{i\sigma} - i \sum_{i} \varphi_{i}
[\sum_{\sigma}{f}^{\dagger}_{i\sigma}{f}_{i\sigma} - 1] \nn && -
t\alpha\sum_{\ij\sigma}(f_{i\sigma}^{\dagger}e^{-ic_{ij}}f_{j\sigma}
+ H. c.) \nn && -
\Delta\sum_{\ij}\varsigma_{ij}e^{-ia_{ij}}(f_{i\uparrow}^{\dagger}f_{j\downarrow}^{\dagger}
- f_{i\downarrow}^{\dagger}f_{j\uparrow}^{\dagger}) - H. c. , \nn
&& L_{\theta} = \frac{1}{U}\sum_{i} (\partial_{\tau}\theta_{i} -
\varphi_{i})^{2} - t\beta
\sum_{\ij}(e^{i\theta_{i}}e^{ic_{ij}}e^{-i\theta_{j}} + H. c.) ,
\nn \eqa where the hopping parameters are represented as
$\alpha_{ij} = \alpha e^{ic_{ij}}$ and $\beta_{ij} = \beta
e^{ic_{ij}}$ with amplitudes $\alpha$, $\beta$ and phase $c_{ij}$,
and $d-wave$ pairing-symmetry is assumed in $\varsigma_{ij}$ of
$|\Delta_{ij}| = \Delta \varsigma_{ij}$ with the gauge
transformation $a_{ij} \rightarrow a_{ij} + \theta_{i} +
\theta_{j}$. The unidentified hopping and pairing amplitudes can
be determined self-consistently in the saddle-point analysis,
$\alpha = |\langle e^{-i\theta_{i}}e^{i\theta_{j}} +
H.c.\rangle|$, $\beta =
|\langle\sum_{\sigma}f_{i\sigma}^{\dagger}f_{j\sigma} +
H.c.\rangle|$ and $\Delta \varsigma_{ij} = |\langle
f_{i\uparrow}^{\dagger}f_{j\downarrow}^{\dagger} -
f_{i\downarrow}^{\dagger}f_{j\uparrow}^{\dagger}\rangle|$.

This effective Lagrangian is quite appealing. Starting from the
BCS-Hubbard effective model, we extract dynamics of collective
charge fluctuations from the Hubbard-$U$ term. As a result, the
fermion sector corresponds to the BCS description with a
renormalized bandwidth $D\alpha$ while dynamics of collective
density excitations is described by the boson-Hubbard-type model
with an effective chemical potential in the saddle-point
approximation ($c_{ij} = 0$). Remember that the boson-Hubbard
model is the prototype for the quantum phase transition associated
with charge fluctuations, describing the genuine Mott transition
without symmetry breaking via condensation. The quantum transition
indeed occurs when $D\beta/U \sim 1$ with the half bandwidth $D$,
where $D\beta$ is an effective bandwidth for the boson field. One
can show that the hopping parameter $\beta$ decreases as $U$
increases, resulting in the bandwidth-control Mott-Hubbard
transition. Actually, the previous slave-rotor studies showed this
transition without superconductivity ($\Delta = 0$) at half
filling.\cite{Florens,LeeLee,Kim_SR_disorder}

However, this treatment does not take into account phase
fluctuations $e^{-ia_{ij}}$ of fermion pairs carefully although
local density fluctuations are governed by U(1) rotor excitations
$e^{i\theta_{i}}$. In other words, the U(1) slave-rotor
representation cannot give any condition for pairing excitations.
Considering the mathematical structure of Eq. (3), one can find
two kinds of gauge excitations, $c_{ij}$ and $a_{ij}$ associated
with local density fluctuations and pairing excitations.
Density-gauge excitations ($c_{ij}$) can be controlled by U(1)
rotor excitations $e^{i\theta_{i}}$. When local density
fluctuations are suppressed in the large $U$ limit, boson
excitations are gapped (Mott insulator), causing gapless
density-gauge fluctuations. On the other hand, when local density
fluctuations become strong in the small $U$ limit, the U(1) rotor
variables get condensed, making density-gauge fluctuations massive
due to the Anderson-Higgs mechanism. However, there are no such
boson excitations corresponding to the U(1) density-rotor variable
for pair-gauge fluctuations ($a_{ij}$).

This discussion motivates us to introduce new boson excitations
associated with local pairing fluctuations, allowing us to control
phase fluctuations of pairing excitations (pair-gauge
fluctuations). Assume the presence of such boson excitations. We
can estimate that both density and pairing fluctuations would be
gapped due to the density-phase uncertainty in the large $U$
limit. On the other hand, both collective excitations become
condensed in the small $U$ limit. What happens near the Mott
critical point? It should be noted that local pair excitations are
different from local density fluctuations. Thus, there is no
reason for both excitations to be condensed at the same time. Can
there exist an intermediate phase, where only one kind of boson
excitations is condensed. If possible, what is the nature of this
intermediate phase?

In this paper we extend the U(1) slave-rotor formulation, allowing
local pairing fluctuations. Then, the collective boson field is
expressed as an SU(2) matrix field $U_{i\sigma\sigma'} =
\left( \begin{array}{cc} z_{i\uparrow}  & - z_{i\downarrow}^{\dagger} \\
z_{i\downarrow} & z_{i\uparrow}^{\dagger}
\end{array} \right)$, involved with both
density ($z_{i\uparrow}$) and pairing ($z_{i\downarrow}$)
fluctuations.\cite{Slave_boson} This leads us to construct an
SU(2) slave-rotor gauge theory of the Hubbard model in terms of
the collective density- and pairing-fluctuation modes interacting
with gapless fermion excitations via SU(2) gauge
fluctuations.\cite{Kim_SU2_SRGT} Compared with the U(1)
slave-rotor gauge theory [Eq. (3)], $z_{i\uparrow}$ corresponds to
$e^{i\theta_{i}}$ roughly speaking while $z_{i\downarrow}$ is
newly introduced to control pairing-gauge fluctuations (phase
fluctuations of electron pairs). Since the SU(2) slave-rotor gauge
theory admits two kinds of collective bosons, an intermediate
phase can be naturally allowed between the spin liquid Mott
insulator ($\langle{z}_{i\uparrow}\rangle = 0$,
$\langle{z}_{i\downarrow}\rangle = 0$) and Fermi liquid metal
($\langle{z}_{i\uparrow}\rangle \not= 0$,
$\langle{z}_{i\downarrow}\rangle \not= 0$), characterized by
softening of the density-fluctuation modes
($\langle{z}_{i\uparrow}\rangle \not= 0$) as the Fermi liquid, but
gapping of the pair-excitation modes
($\langle{z}_{i\downarrow}\rangle = 0$) as the spin liquid.
Remember that the U(1) slave-rotor gauge theory contains only the
density-fluctuation modes ($z_{i\uparrow}$), thus allowing the two
phases only, the spin liquid Mott insulator
($\langle{z}_{i\uparrow}\rangle = 0$) and Fermi liquid metal
($\langle{z}_{i\uparrow}\rangle \not= 0$).

Performing a mean-field analysis based on an SU(2) slave-rotor
effective action, we find the intermediate phase indeed away from
half filling, described by partial freezing of charge fluctuations
($\langle{z}_{i\uparrow}\rangle \not= 0$,
$\langle{z}_{i\downarrow}\rangle = 0$) near the Mott-Hubbard
critical point. We reveal that this intermediate phase originates
from emergence of a pseudospin-dependent chemical potential due to
hole doping that causes SU(2) pseudospin symmetry breaking to
discriminate the local pair-excitation modes ($z_{i\downarrow}$)
from the local density-fluctuation modes ($z_{i\uparrow}$).
Condensation of density-fluctuation modes makes density-gauge
excitations gapped due to the Anderson-Higgs mechanism. On the
other hand, pairing-gauge fluctuations remain gapless because
pairing-excitation modes are gapped. As a result, we find an
effective U(1) gauge theory for this intermediate phase, where
phase fluctuations of fermion pairs are described by pairing-gauge
fluctuations. We discuss physics of this non-superconducting
phase, identified with a non-Fermi liquid metal due to the
presence of phase-incoherent pairs.

\section{SU(2) slave-rotor theory of the Hubbard model}

\subsection{SU(2) slave-rotor representation in the path integral formulation}

We start from the Hubbard Hamiltonian \bqa && H = -
t\sum_{\ij\sigma}(c_{i\sigma}^{\dagger}c_{j\sigma} + H.c.) +
\frac{3u}{2}\sum_{i}c_{i\uparrow}^{\dagger}c_{i\uparrow}c_{i\downarrow}^{\dagger}c_{i\downarrow}
, \eqa where $u/t$ is a parameter determining various phases of
this model with hole concentration. The local interaction term can
be decomposed into the pairing and density channels in the
following way \bqa &&
\frac{3u}{2}\sum_{i}c_{i\uparrow}^{\dagger}c_{i\uparrow}c_{i\downarrow}^{\dagger}c_{i\downarrow}
=
\frac{u}{2}\sum_{i}c_{i\uparrow}^{\dagger}c_{i\downarrow}^{\dagger}
c_{i\downarrow}c_{i\uparrow} \nn && +
\frac{u}{2}\sum_{i}\Bigl(\sum_{\sigma}c_{i\sigma}^{\dagger}c_{i\sigma}-1\Bigr)^{2}
+
\frac{u}{2}\Bigl(\sum_{\sigma}c_{i\sigma}^{\dagger}c_{i\sigma}-1\Bigr)
. \nonumber \eqa Performing the HS transformation for the pairing-
and density-interaction channels, we find an effective Lagrangian
in the Nambu-spinor representation \bqa &&  Z = \int{D[\psi_{i},
\Phi^{R}_{i}, \Phi_{i}^{I},
\varphi_{i}]}e^{-\int_{0}^{\beta}{d\tau} L} , \nn && L =
\sum_{i}\psi_{i}^{\dagger}(\partial_{\tau}\mathbf{I} -
\mu\tau_{3})\psi_{i} -
t\sum_{\ij}(\psi_{i}^{\dagger}\tau_{3}\psi_{j} + H.c.) \nn && -
i\sum_{i}(\Phi^{R}_{i}\psi_{i}^{\dagger}\tau_{1}\psi_{i} +
\Phi^{I}_{i}\psi_{i}^{\dagger}\tau_{2}\psi_{i} +
\varphi_{i}\psi_{i}^{\dagger}\tau_{3}\psi_{i}) \nn && +
\frac{1}{2u}\sum_{i}(\Phi^{R2}_{i} + \Phi^{I2}_{i} +
\varphi_{i}^{2}) . \eqa Here $\psi_{i}$ is the Nambu spinor, given
by $\psi_{i}= \left(\begin{array}{c} c_{i\uparrow} \\
c_{i\downarrow}^{\dagger} \end{array} \right)$.  $\Phi^{R}_{i}$
and $\Phi^{I}_{i}$ are the real and imaginary parts of the on-site
pairing-order parameter respectively, and $\varphi_{i}$ is an
effective density-potential. $\mu$ is an electron chemical
potential, renormalized by local interactions as $\mu = \mu_{b} -
u/2$, where $\mu_{b}$ is the bare chemical potential. Introducing
a pseudospin vector $\vec{\Omega}_{i} \equiv (\Phi_{i}^{R},
\Phi_{i}^{I}, \varphi_{i})$, one can express Eq. (5) in a compact
form \bqa && Z = \int{D[\psi_{i},
\vec{\Omega}_{i}]}e^{-\int_{0}^{\beta}{d\tau} L} , \nn && L =
\sum_{i}\psi_{i}^{\dagger}(\partial_{\tau}\mathbf{I} -
\mu\tau_{3})\psi_{i} -
t\sum_{\ij}(\psi_{i}^{\dagger}\tau_{3}\psi_{j} + H.c.) \nn && -
i\sum_{i}\psi_{i}^{\dagger}(\vec{\Omega}_{i}\cdot\vec{\tau})\psi_{i}
+
\frac{1}{4u}\sum_{i}\mathbf{tr}(\vec{\Omega}_{i}\cdot\vec{\tau})^{2}
. \eqa Integrating over the pseudospin field $\vec{\Omega}_{i}$,
Eq. (6) recovers the Hubbard model Eq. (4). Note that the U(1)
slave-rotor representation allows only the density-interaction
channel including the $\tau_{3}$ matrix.

As discussed in the introduction, we disintegrate bare electrons
into collective excitations and renormalized electrons in the
following way \bqa \psi_{i} & = &
e^{-i\phi_{1i}\tau_{1}-i\phi_{2i}\tau_{2}-i\phi_{3i}\tau_{3}}\eta_{i}
\equiv U_{i}^{\dagger}\eta_{i} , \eqa  where the two component
spinor $\eta_{i} = \left( \begin{array}{c} \eta_{i+} \\
\eta_{i-}^{\dagger} \end{array} \right)$ can be considered to
express renormalized electrons, and the SU(2) matrix field $U_{i}
= \exp[i\sum_{k=1}^{3}\phi_{ki}\tau_{k}]$ collective bosons. Here
$\exp[i\phi_{1i}\tau_{1}]$ ($\exp[i\phi_{2i}\tau_{2}]$) can be
interpreted as a creation operator of an electron pair since it
mixes a particle with a hole. If one takes only the $\phi_{1i}$
phase field in the $U_{i}$ matrix field, i.e., $U_{i} =
e^{i\phi_{1i}\tau_{1}}$, he finds $\eta_{i+} =
\cos\phi_{i}c_{i\uparrow} +
i\sin\phi_{i}c_{i\downarrow}^{\dagger}$ and $\eta_{i-}^{\dagger} =
i\sin\phi_{i}c_{i\uparrow} +
\cos\phi_{i}c_{i\downarrow}^{\dagger}$ from $\eta_{i} =
e^{i\phi_{1i}\tau_{1}}\psi_{i}$. Considering the $\phi_{2i}$ phase
field, one obtains $\eta_{i+} = \cos\phi_{i}c_{i\uparrow} +
\sin\phi_{i}c_{i\downarrow}^{\dagger}$ and $\eta_{i-}^{\dagger} =
-\sin\phi_{i}c_{i\uparrow} +
\cos\phi_{i}c_{i\downarrow}^{\dagger}$ in the same way. On the
other hand, $\exp[i\phi_{3i}\tau_{3}]$ is identified with a
creation operator of an electron charge, corresponding to the
rotor variable in the U(1) slave-rotor representation.

Inserting the SU(2) slave-rotor decomposition Eq. (7) into the
effective Lagrangian Eq. (6), we obtain \bqa && Z =
\int{D[\eta_{i}, U_{i}, \vec{\Omega}_{i}]}
e^{-\int_{0}^{\beta}{d\tau} L} , \nn && L =
\sum_{i}\eta_{i}^{\dagger}(\partial_{\tau}\mathbf{I} +
U_{i}\partial_{\tau}U_{i}^{\dagger}-
\mu{U}_{i}\tau_{3}U_{i}^{\dagger})\eta_{i} \nn && -
t\sum_{\ij}(\eta_{i}^{\dagger}U_{i}\tau_{3}U_{j}^{\dagger}\eta_{j}
+ H.c.) \nn && -
i\sum_{i}\eta_{i}^{\dagger}U_{i}(\vec{\Omega}_{i}\cdot\vec{\tau})U_{i}^{\dagger}\eta_{i}
+
\frac{1}{4u}\sum_{i}\mathbf{tr}(\vec{\Omega}_{i}\cdot\vec{\tau})^{2}
. \eqa Performing the unitary transformation
$\vec{\Omega}_{i}\cdot\vec{\tau} \rightarrow
U_{i}^{\dagger}(\vec{\Omega}_{i}\cdot\vec{\tau})U_{i}$ that makes
the integration-measure invariant, Eq. (8) reads \bqa && Z =
\int{D[\eta_{i}, U_{i}, \vec{\Omega}_{i}]}
e^{-\int_{0}^{\beta}{d\tau} L} , \nn && L =
\sum_{i}\eta_{i}^{\dagger}(\partial_{\tau}\mathbf{I} +
U_{i}\partial_{\tau}U_{i}^{\dagger}-
\mu{U}_{i}\tau_{3}U_{i}^{\dagger})\eta_{i} \nn && -
t\sum_{\ij}(\eta_{i}^{\dagger}U_{i}\tau_{3}U_{j}^{\dagger}\eta_{j}
+ H.c.) \nn && -
i\sum_{i}\eta_{i}^{\dagger}(\vec{\Omega}_{i}\cdot\vec{\tau})\eta_{i}
+
\frac{1}{4u}\sum_{i}\mathbf{tr}(\vec{\Omega}_{i}\cdot\vec{\tau})^{2}
. \eqa Shifting $\vec{\Omega}_{i}\cdot\vec{\tau}$ as
$\vec{\Omega}_{i}\cdot\vec{\tau} - i
U_{i}\partial_{\tau}U_{i}^{\dagger} +
i\mu{U}_{i}\tau_{3}U_{i}^{\dagger}$ that also makes the
integration-measure invariant, we find the SU(2) slave-rotor
representation of the Hubbard model as an extended version of its
U(1) slave-rotor representation \bqa && Z = \int{D[\eta_{i},
U_{i}, \vec{\Omega}_{i}]} e^{-\int_{0}^{\beta}{d\tau} L} , \nn &&
L = \sum_{i}\eta_{i}^{\dagger}(\partial_{\tau}\mathbf{I} -
i\vec{\Omega}_{i}\cdot\vec{\tau})\eta_{i} -
t\sum_{\ij}(\eta_{i}^{\dagger}U_{i}\tau_{3}U_{j}^{\dagger}\eta_{j}
+ H.c.) \nn && +
\frac{1}{4u}\sum_{i}\mathbf{tr}(-iU_{i}\partial_{\tau}U_{i}^{\dagger}
+ \vec{\Omega}_{i}\cdot\vec{\tau} +
i\mu{U}_{i}\tau_{3}U_{i}^{\dagger})^{2} . \eqa

\subsection{Relation between the path integral formulation and
canonical quantization}

One cautious person may suspect the above derivation because any
decomposition method should always impose its associated
constraint, but the path integral derivation does not seem to
include an appropriate constraint. However, this guess is not
correct because the above derivation imposes its constraint
indeed.\cite{Kim_SR_disorder,Kim_SR_doping}

We start from the easy-axis anisotropy with $\vec{\Omega}_{i} =
(0, 0, \varphi_{i})$ and $U_{i} = e^{i\phi_{3}\tau_{3}}$ in order
to check whether Eq. (10) recovers the U(1) slave-rotor Lagrangian
[Eq. (3) with $\Delta = 0$] in this limit. Inserting this
easy-axis representation into Eq. (10), and performing the HS
transformation for the last time-fluctuation term, we obtain \bqa
&& Z = \int{D[\eta_{i},\phi_{3i},\varphi_{i},L_{i}]}
\exp\Bigl[-\int_{0}^{\beta}{d\tau} \Bigl\{
\sum_{i}\eta_{i}^{\dagger}\partial_{\tau}\eta_{i} \nn && -
t\sum_{\ij}(\eta_{i}^{\dagger}e^{i\phi_{3i}\tau_{3}}\tau_{3}e^{-i\phi_{3j}\tau_{3}}\eta_{j}
+ H.c.) \nn && + \sum_{i}\Bigl(\frac{u}{2}L_{i}^{2} -
iL_{i}(\partial_{\tau}\phi_{3i} - i\mu) + i\varphi_{i}(L_{i} -
\eta_{i}^{\dagger}\tau_{3}\eta_{i})\Bigr) \Bigr\} \Bigr] , \nn
\eqa where $\eta_{i}$ and $e^{-i\phi_{3i}\tau_{3}}$ carry spin and
charge degrees of freedom, respectively. In this expression the
electron Hilbert space $|\psi_{i}>$ is decomposed into the direct
product of the boson and fermion Hilbert spaces
${|L_{i}>}\bigotimes|\eta_{i}>$ according to the decomposition
$\psi_{i} = e^{-i\phi_{3i}\tau_{3}}\eta_{i}$, where $L_{i}$
represents an electron density at site $i$. Since this
decomposition enlarges the original electron Hilbert space, the
U(1) slave-rotor constraint $L_{i} =
\eta_{i}^{\dagger}\tau_{3}\eta_{i}$ is introduced to reduce the
enlarged Hilbert space into the original electron one. Then,
$e^{-i\phi_{3i}\tau_{3}}$ is identified with an annihilation
operator of an electron charge in the Nambu-spinor representation
owing to the constraint $L_{i} =
\eta_{i}^{\dagger}\tau_{3}\eta_{i}$ and the canonical relation
$[L_{i}, \phi_{3j}] = -i\delta_{ij}$ imposed by $-
iL_{i}\partial_{\tau}\phi_{3i}$. Actually, Eq. (11) is the
starting point in the canonical quantization of the U(1)
slave-rotor representation for the Hubbard model, Eq. (3) with
$\Delta = 0$ as discussed in the
introduction.\cite{Florens,Kim_SR_disorder,Kim_SR_doping} In this
respect the above expression connects the canonical approach with
the path integral representation.

In the same way as the above, performing the HS transformation for
the last time-fluctuation term in Eq. (10) \bqa &&
\frac{1}{4u}\sum_{i}\mathbf{tr}(-iU_{i}\partial_{\tau}U_{i}^{\dagger}
+ \vec{\Omega}_{i}\cdot\vec{\tau} +
i\mu{U}_{i}\tau_{3}U_{i}^{\dagger})^{2} \nn && \rightarrow
\sum_{i}\mathbf{tr} \Bigl\{ u M_{i}^{2} +
iM_{i}(-iU_{i}\partial_{\tau}U_{i}^{\dagger} +
\vec{\Omega}_{i}\cdot\vec{\tau} +
i\mu{U}_{i}\tau_{3}U_{i}^{\dagger}) \Bigr\} , \nonumber \eqa we
obtain the following expression for the SU(2) slave-rotor
representation \bqa && L =
\sum_{i}\eta_{i}^{\dagger}\partial_{\tau} \eta_{i} -
t\sum_{\ij}(\eta_{i}^{\dagger}U_{i}\tau_{3}U_{j}^{\dagger}\eta_{j}
+ H.c.) \nn && + \sum_{i} \Bigl[ \mathbf{tr} \Bigl\{ u M_{i}^{2} +
iM_{i}(-iU_{i}\partial_{\tau}U_{i}^{\dagger}+
i\mu{U}_{i}\tau_{3}U_{i}^{\dagger})\Bigr\} \nn && + i{\vec
\Omega}_{i}\cdot(\mathbf{tr}[M_{i}{\vec \tau}] -
\eta_{i}^{\dagger}\vec{\tau}\eta_{i}) \Bigr] , \eqa where $M_{i}$
is an SU(2) pseudospin matrix. In this SU(2) case the electron
Hilbert space is represented as $|\psi_{i}> =
{|M_{i}>}\bigotimes|\eta_{i}>$ according to the decomposition
$\psi_{i} = U_{i}^{\dagger}\eta_{i}$. The Lagrange multiplier
field $\vec{\Omega}_{i}$ confirms the local constraint
$\mathbf{tr}[M_{i}{\vec \tau}] =
\eta_{i}^{\dagger}\vec{\tau}\eta_{i}$. Using $M_{i} =
\vec{N}_{i}\cdot\vec{\tau}$ where $\vec{N}_{i}$ corresponds to the
pseudospin-density-wave order parameter, one can show that this is
nothing but the standard relation $\vec{N}_{i} =
\frac{1}{2}\eta_{i}^{\dagger}\vec{\tau}\eta_{i}$ as a natural
extension of the U(1) slave-rotor constraint. $U_{i}^{\dagger}$ is
identified with an annihilation operator of an electron pseudospin
in the same way as the U(1) case.

\subsection{SU(2) slave-rotor gauge theory of the Hubbard
model}

Using the HS transformation for the correlated hopping term in Eq.
(10) \bqa && -
t(\eta_{i\alpha}^{\dagger}U_{i\alpha\beta}\tau_{3\beta\gamma}U_{j\gamma\delta}^{\dagger}\eta_{j\delta}
+ H.c.) \nn && \rightarrow
t\Bigl[F_{ij\alpha\delta}E_{ij\delta\alpha}^{\dagger} +
E_{ij\alpha\delta}F_{ij\delta\alpha}^{\dagger} \nn && -
(\eta_{i\alpha}^{\dagger}F_{ij\alpha\delta}\eta_{j\delta} +
U_{i\alpha\beta}\tau_{3\beta\gamma}U_{j\gamma\delta}^{\dagger}E_{ij\delta\alpha}^{\dagger})
- H.c. \Bigr] , \nonumber \eqa we find an effective Lagrangian of
the Hubbard model \bqa && Z = \int{D[\eta_{i}, U_{i},
\vec{\Omega}_{i}, E_{ij}, F_{ij}]}e^{-\int_{0}^{\beta}{d\tau} L} ,
\nn && L = L_{0} + L_{\eta} + L_{U} , \nn && L_{0} =
t\sum_{\ij}\mathbf{tr}(F_{ij}E_{ij}^{\dagger} +
E_{ij}F_{ij}^{\dagger}) , \nn && L_{\eta} =
\sum_{i}\eta_{i}^{\dagger}(\partial_{\tau}\mathbf{I} -
i\vec{\Omega}_{i}\cdot\vec{\tau})\eta_{i} -
t\sum_{\ij}(\eta_{i}^{\dagger}F_{ij}\eta_{j} + H.c.) , \nn &&
L_{U} =
\frac{1}{4u}\sum_{i}\mathbf{tr}(-iU_{i}\partial_{\tau}U_{i}^{\dagger}
+ \vec{\Omega}_{i}\cdot\vec{\tau} +
i\mu{U}_{i}\tau_{3}U_{i}^{\dagger})^{2} \nn && -
t\sum_{\ij}\mathbf{tr}(U_{j}^{\dagger}E_{ij}^{\dagger}U_{i}\tau_{3}
+ H.c.) , \eqa where $E_{ij}$ and $F_{ij}$ are HS matrix fields
associated with hopping of $\eta_{i}$ fermions and $U_{i}$ bosons,
respectively.

We make an ansatz for the hopping matrix fields as \bqa && E_{ij}
\approx E\exp[i\vec{a}_{ij}\cdot\vec{\tau}]\tau_{3} , ~~~ F_{ij}
\approx F\exp[i\vec{a}_{ij}\cdot\vec{\tau}]\tau_{3} , \eqa where
$E$ and $F$ are longitudinal modes (amplitudes) of the hopping
parameters, and $\vec{a}_{ij}$ their transverse modes (phase
fluctuations), considered to be spatial components of SU(2) gauge
fields. The reason why we introduce the $\tau_{3}$ matrix is that
the fermion sector $L_{\eta}$ should recover the original electron
Lagrangian [Eq. (6)] as the slave-rotor representation does.

Inserting Eq. (14) into Eq. (13), we reach the SU(2) slave-rotor
gauge theory of the Hubbard model for the metal-insulator
transition \bqa && Z = \int{D[\eta_{i}, U_{i}, \vec{\Omega}_{i},
\vec{a}_{ij}]}e^{-\int_{0}^{\beta}{d\tau} L} , \nn && L = L_{\eta}
+ L_{U} + 4t\sum_{\ij} EF , \nn && L_{\eta} =
\sum_{i}\eta_{i}^{\dagger}(\partial_{\tau}\mathbf{I} -
i\vec{\Omega}_{i}\cdot\vec{\tau})\eta_{i} -
tF\sum_{\ij}(\eta_{i}^{\dagger}e^{i\vec{a}_{ij}\cdot\vec{\tau}}\tau_{3}\eta_{j}
+ H.c.) , \nn && L_{U} =
\frac{1}{4u}\sum_{i}\mathbf{tr}(-iU_{i}\partial_{\tau}U_{i}^{\dagger}
+ \vec{\Omega}_{i}\cdot\vec{\tau} +
i\mu{U}_{i}\tau_{3}U_{i}^{\dagger})^{2} \nn && -
tE\sum_{\ij}\mathbf{tr}(U_{j}^{\dagger}\tau_{3}e^{-i\vec{a}_{ij}\cdot\vec{\tau}}U_{i}\tau_{3}
+ H.c.) , \eqa where the unknown parameters such as the amplitudes
of the hopping parameters $E$, $F$ and the SU(2) pseudospin order
parameter $\vec{\Omega}_{i}$ should be determined by the following
self-consistent mean-field equations \bqa && 4E =
\langle\eta_{i}^{\dagger}\tau_{3}\eta_{j} + H.c. \rangle , ~~~ 4F
= \langle\mathbf{tr}({U}_{j}^{\dagger}\tau_{3}U_{i}\tau_{3} +
H.c.)\rangle , \nn && {\vec \Omega}_{i} =
iu\langle\eta_{i}^{\dagger}\vec{\tau}\eta_{i}\rangle -
\frac{1}{2}\langle\mathbf{tr}[\vec{\tau}(-iU_{i}\partial_{\tau}U_{i}^{\dagger}
+ i\mu{U}_{i}\tau_{3}U_{i}^{\dagger})]\rangle , \eqa as will be
discussed later.

Compared with the U(1) slave-rotor effective action Eq. (3), the
SU(2) slave-rotor effective action Eq. (15) exhibits more fruitful
physics. Dynamics of collective charge fluctuations can be
extracted from interacting electrons in the strong coupling
approach as the U(1) case, but the structure of their dynamics is
much richer. The effective boson action is represented as the
nonlinear $\sigma$ model-type (its CP$^1$ representation) with an
effective chemical potential (the time component of an SU(2) gauge
field) in the saddle-point approximation ignoring SU(2) gauge
fluctuations while that in the U(1) slave-rotor representation is
governed by the boson-Hubbard model-type. The presence of
additional collective charge fluctuations opens the possibility of
new phases, as will be discussed in this paper.

Considering the fermion sector of the SU(2) slave-rotor
representation, one cautious person can find the possibility of
superconductivity in the Hubbard model. See the kinetic energy
term of the fermion sector in Eq. (15). Because the SU(2) gauge
matrix $W_{ij} \equiv \exp[- i\vec{a}_{ij}\cdot\vec{\tau}]$ has
nonzero off-diagonal components, fermion pairing as the seed for
superconductivity is naturally allowed. Local repulsive
interactions (the Hubbard-$U$ term) appear to be local density and
pairing fluctuations of the SU(2) matrix field $U_{i}$, governed
by the nonlinear $\sigma$ model-type. These charge fluctuations
generate SU(2) gauge fluctuations and couple to them, one of which
corresponds to phase fluctuations of fermion pairs, observed in
the kinetic energy term of the fermion sector. Since the SU(2)
gauge fluctuations are controlled via the collective charge
fluctuations, fermion-pairing fluctuations can be managed by the
collective boson-dynamics, varying the coupling constant $u/t$ and
hole concentration $\delta$. In summary, the SU(2) slave-rotor
representation not only reveals the possibility of
superconductivity in the Hubbard model, but also controls phase
fluctuations of fermion pairs.

\subsection{U(1) charge-rotor gauge theory of the Hubbard model}

Since the structure of the SU(2) slave-rotor action [Eq. (15)] is
complex to analyze, it is necessary to consider its easy-axis
limit $\vec{\Omega}_{i}\cdot\vec{\tau} \equiv
\varphi_{i}\tau_{3}$, $U_{i} \equiv \exp[i\phi_{3i}\tau_{3}]$, and
${\vec a}_{ij}\cdot{\vec \tau} \equiv a_{3ij}\tau_{3}$. Then, Eq.
(15) is reduced to the effective U(1) gauge Lagrangian in the
slave-rotor
representation\cite{Florens,LeeLee,Kim_SR_disorder,Kim_SU2_SRGT,Kim_SR_doping}
\bqa && L_{\eta} =
\sum_{i}\eta_{i}^{\dagger}(\partial_{\tau}\mathbf{I} -
i\varphi_{i}\tau_{3})\eta_{i} \nn && -
tF\sum_{\ij}(\eta_{i}^{\dagger}e^{i{a}_{3ij}\tau_{3}}\tau_{3}\eta_{j}
+ H.c.) , \nn && L_{\phi} =
\frac{1}{2u}\sum_{i}(\partial_{\tau}\phi_{3i} - \varphi_{i} -
i\mu)^{2} \nn && - 2tE\sum_{\ij}\cos(\phi_{3j} - \phi_{3i} -
a_{3ij}) . \eqa Based on this slave-rotor effective Lagrangian,
Florens and Georges performed a saddle-point analysis at half
filling, and found a coherent-incoherent transition of the
$\phi_{3i}$ fields, identifying this transition with the
Mott-Hubbard transition from a spin liquid Mott insulator to a
Fermi liquid metal.\cite{Florens} For our later discussion we
perform the mean-field analysis of the U(1) slave-rotor theory.

For the saddle-point analysis one can resort to the large $N$
generalization for the boson sector replacing $e^{i\phi_{3i}}$
with $X_{i\sigma}$ \bqa && L_{X} =
\frac{1}{2u}\sum_{i\sigma}(\partial_{\tau}X_{i\sigma}^{\dagger} -
[\mu - \varphi_{i}] X_{i\sigma}^{\dagger})(\partial_{\tau}
X_{i\sigma} + [\mu - \varphi_{i}]X_{i\sigma}) \nn && -
tE\sum_{ij\sigma} X_{j\sigma}^{\dagger}X_{i\sigma} +
i\sum_{i}\lambda_{i}(\sum_{\sigma}|X_{i\sigma}|^{2} - 1) , \eqa
where $\lambda_{i}$ is a Lagrange multiplier field to impose the
uni-modular constraint for the rotor field $X_{i\sigma}$, and
$\varphi_{i}$ is substituted for $i\varphi_{i}$. Based on the
mean-field ansatz of $\varphi_{i} = \varphi_{0}$, $i\lambda_{i} =
\lambda$, and $a_{3ij} = 0$, we find the mean-field free energy
\bqa && F_{MF} = - \frac{1}{\beta}\sum_{k,\omega}\ln[(i\omega)^{2}
- (\varphi_{0} - F\epsilon_{k}^{\eta})^{2}] \nn && +
\frac{1}{\beta} \sum_{k,\nu}\sum_{\sigma}\ln[-\frac{(i\nu + \mu -
\varphi_{0} )^{2}}{2u} + \lambda + E\epsilon_{k}^{X}] \nn && +
\sum_{k}(DEF - \mu\delta - \lambda) , \eqa where $\omega$ ($\nu$)
is the Matsubara frequency for fermions (bosons) with inverse
temperature $\beta = 1/T$, and $\epsilon_{k}^{\eta(X)} = -2 t
(\cos k_x + \cos k_y)$ is the fermion (boson) dispersion that
originates from the electron dispersion. $\delta$ is hole
concentration.

Minimizing the free energy with respect to $E$, $F$,
$\varphi_{0}$, $\lambda$, and $\mu$, we find the self-consistent
saddle-point equations for the mean-field parameters \bqa && DE =
\int_{-D}^{D}d\epsilon{D}(\epsilon)\frac{1}{\beta}\sum_{\omega}\frac{2(\varphi_{0}
- F\epsilon)\epsilon }{(i\omega)^{2} - (\varphi_{0} -
F\epsilon)^{2} } , \nn &&  DF = -
\int_{-D}^{D}d\epsilon{D}(\epsilon)\frac{1}{\beta}\sum_{\nu\sigma}\frac{\epsilon}{-\frac{(i\nu
+ \mu - \varphi_{0})^{2}}{2u} + \lambda + E\epsilon} , \nn &&
\int_{-D}^{D}d\epsilon{D}(\epsilon)\frac{1}{\beta}\sum_{\omega}\frac{2(\varphi_{0}
- F\epsilon)}{(i\omega)^{2} - (\varphi_{0} - F\epsilon)^{2} } \nn
&& = -
\int_{-D}^{D}d\epsilon{D}(\epsilon)\frac{1}{\beta}\sum_{\nu\sigma}\frac{\frac{i\nu
+ \mu - \varphi_{0}}{u}}{-\frac{(i\nu +\mu - \varphi_{0})^{2}}{2u}
+ \lambda + E\epsilon} , \nn &&  1 =
\int_{-D}^{D}d\epsilon{D}(\epsilon)\frac{1}{\beta}\sum_{\nu\sigma}
\frac{1}{-\frac{(i\nu + \mu - \varphi_{0})^{2}}{2u} + \lambda +
E\epsilon} , \nn &&  - \delta =
\int_{-D}^{D}d\epsilon{D}(\epsilon)\frac{1}{\beta}\sum_{\nu\sigma}\frac{\frac{i\nu
+ \mu - \varphi_{0}}{u}}{-\frac{(i\nu + \mu -
\varphi_{0})^{2}}{2u} + \lambda + E\epsilon} . \eqa Here
$\sum_{k}$ is replaced with $\int_{-D}^{D}d\epsilon D(\epsilon)$,
where $D$ is the half bandwidth and $D(\epsilon)$ is the density
of states for an electron band.

Performing the Matsubara frequency summation and energy
integration with the constant density of states $D(\epsilon) =
1/(2D)$ in Eq. (20), we obtain the algebraic equations for the
mean-field parameters $E$, $F$, $\varphi_{0}$, $\lambda$, and
$\mu$ at zero temperature \bqa && E = \frac{1}{2}\Bigl[1 -
\Bigl(\frac{\varphi_{0}}{DF}\Bigr)^{2}\Bigr] , \nn && F =
\frac{1}{3(DE)^2}\Bigl[ (2\lambda-DE)\sqrt{2u(\lambda+DE)} \nn &&
- (2\lambda-E\epsilon_{-})\sqrt{2u(\lambda+E\epsilon_{-})} \Bigr]
, \nn && - \frac{\varphi_{0}}{DF} = 1 - \frac{\lambda}{DE} +
\frac{1}{DE}\frac{(\mu-\varphi_{0})^{2}}{2u} , \nn && 1 =
\frac{\sqrt{2u(\lambda + DE)} -
\sqrt{2u(\lambda+E\epsilon_{-})}}{DE} , \nn && \delta = 1 -
\frac{\lambda}{DE} + \frac{1}{DE}\frac{(\mu-\varphi_{0})^{2}}{2u}
, \eqa where $\epsilon_{-}$ is given by $\epsilon_{-} =
\frac{1}{E}\Bigl[\frac{(\mu-\varphi_{0})^{2}}{2u} -
\lambda\Bigr]$. From the third and fifth equations we find
$\varphi_{0} = -DF \delta$ and $\mu = - DF \delta -
\sqrt{2u[\lambda - DE(1- \delta)]}$. Inserting these expressions
into the above equations, we obtain \bqa && E =
\frac{1-\delta^2}{2} , \nn && F = \frac{1}{3(DE)^2}\Bigl[
(2\lambda-DE)\sqrt{2u(\lambda+DE)} \nn && -
[2\lambda+DE(1-\delta)]\sqrt{2u[\lambda - DE(1- \delta)]}\Bigr] ,
\nn && 1 = \frac{\sqrt{2u(\lambda + DE)} - \sqrt{2u[\lambda -
DE(1- \delta)]}}{DE} . \eqa Condensation of the $\phi_{3i}$ bosons
occurs when their excitation gap closes, given by
$\mu_{c}-\varphi_{0c} = 0$ or $\lambda_{c} - DE_{c}(1-\delta) = 0$
that determines the Mott-Hubbard critical point \bqa &&
\frac{u_c}{D} = \frac{1}{4} \Bigl( \frac{1-\delta^2}{2-\delta}
\Bigr) \eqa with $F_{c} = (1-2\delta)/3$ in the mean-field
approximation. This means that the spin liquid Mott insulator
($\langle{e}^{i\phi_{3i}}\rangle = 0$) appears in $u/D > u_c/D$,
and the Fermi liquid metal ($\langle{e}^{i\phi_{3i}}\rangle \not=
0$) arises in $u/D < u_c/D$. It is important to notice that since
the slave-rotor decomposition is meaningful in the case of $E > 0$
and $F > 0$, the critical value of $F_{c} = (1-2\delta)/3$
indicates that this representation is valid when hole
concentration is relatively small, here $\delta < 1/2$.

\subsection{Spin liquid Mott insulator away from half filling}

One would be surprised at this result that the slave-rotor theory
allows the spin liquid Mott insulator away from half filling, in
contrast with the common belief that hole doping to a Mott
insulator is expected to cause a metal. Hole doping to a Mott
insulator shifts the chemical potential from the middle of the
charge gap generated by the energy cost ($u$) for double occupancy
to the top of the lower Hubbard band.\cite{Phillips} Since the
density of states at the top of the lower Hubbard band is nonzero,
the resulting state is expected to be a metallic phase. Actually,
the $t-J$ model studies find a metallic phase at all fillings,
even arbitrarily close to half filling.\cite{tJ_metal}

Recently, Choy and Phillips claimed that this common belief may
not be right in the Hubbard model. Doped Mott insulators can be
insulators.\cite{Phillips} They demonstrated that hole
localization can obtain because the chemical potential lies in a
pseudogap which has vanishing density of states at zero
temperature. The pseudogap in the doped Mott insulator results
from short-range antiferromagnetic correlations corresponding to
the nearest-neighbor singlet-triplet splitting. They showed that
the pseudogap vanishes without the triplet contribution which lies
above the chemical potential, claiming that the presence of the
upper Hubbard band is crucial for the pseudogap. More
fundamentally, they proposed that physics is sensitive to the
order of limits of $u \rightarrow \infty$ and $L \rightarrow
\infty$, where $L$ is the system size. They suggested the
relevance length scale $\xi_{do}$ for the pseudogap, where
$\xi_{do}$ represents the average distance between doubly occupied
sites. They claimed that the order of $u \rightarrow \infty$ and
$L \rightarrow \infty$ results in $\xi_{do} > L$, metallic
transport while $\xi_{do} < L$ and localization obtains in the
reverse order of limits, provided that $n_{h}\xi_{do}^{2} < L$,
$n_{h} = x(L/a)^2$ the number of holes with lattice spacing $a$.
Furthermore, considering the scaling form $Z \sim L^{-(t/u)^{p}}$
with $p > 0$ for the one-hole quasiparticle weight $Z$, they
addressed that the discrepancy between the $t-J$ and Hubbard
results implies the non-commutativity of the order of limits. In
the $t-J$ model (no double occupancy) corresponding to $u
\rightarrow \infty$ and $L \rightarrow \infty$, $Z$ remains
finite. But, $Z$ vanishes in the reverse order of limits (Hubbard
model).

The slave-rotor theory admits double occupancy. In addition, it
was demonstrated that the exchange energy scale $J \sim t^2/u$ is
indeed generated at the level of one-particle properties, where it
cuts the divergence of the effective mass at the Mott transition
driven by $u$.\cite{Florens} Actually, the single-particle
spectral weight is enhanced near the chemical potential with
vanishing density of states at the chemical potential, decreasing
dimensionality from infinite dimensions near the Mott critical
point at half filling, which is associated with the exchange
energy scale.\cite{Florens}

To see the mechanism for the spin liquid Mott insulator away from
half filling in the slave-rotor theory more concretely, we
consider the effective chemical potential in the boson sector of
Eq. (18) carefully. It is important to notice that the effective
chemical potential for the rotor condensation is given by $\mu -
\varphi_{0}$ instead of $\varphi_{0}$ or $\mu$. The point is that
the effective chemical potential $\mu - \varphi_0 = -
\sqrt{2u[\lambda - DE(1- \delta)]}$ away from half filling shows
essentially the same behavior as the chemical potential $\mu = -
\sqrt{2u[\lambda - DE]}$ with $\varphi_{0} = 0$ at half filling
although $\mu = - DF \delta - \sqrt{2u[\lambda - DE(1- \delta)]}$
and $\varphi_0 = - DF \delta$ away from half filling have doping
dependencies proportional to hole
concentration.\cite{Half_filling_Eqs} Considering physics of the
slave-rotor variable, this behavior of the effective chemical
potential can be understood. The U(1) slave-rotor variable raises
or lowers the local density of fermions. Thus, hole concentration
has nothing to do with the density of slave-rotor bosons. This is
in contrast with the slave-boson representation of the $t-J$
model, where hole concentration is exactly the same as the density
of slave-bosons in the saddle-point approximation. In the $t-J$
model the single occupancy constraint of
$f_{i\sigma}^{\dagger}f_{i\sigma} + b_{i}^{\dagger}b_{i} = 1$ with
spinon $f_{i\sigma}$ and holon $b_{i}$ results in
$\langle{f_{i\sigma}^{\dagger}f_{i\sigma}}\rangle = 1 - \delta$
and $\langle{b_{i}^{\dagger}b_{i}}\rangle = \delta$, thus causing
the condensation of holons due to the holon chemical potential as
soon as holes are doped. On the other hand, there is no such
constraint in the slave-rotor representation, thus the density of
slave-rotor bosons does not follow hole concentration. Hole doping
changes the density of fermions only
($\langle\eta_{i}^{\dagger}\tau_{3}\eta_{i}\rangle = - \delta$),
reflected in the doping dependence of the effective chemical
potential $\varphi_0 = - DF \delta$ for fermions. The effect of
hole doping on the boson dynamics is just to modify the bandwidth
from $DE$ to $DE(1-\delta)$. As a result, the slave-rotor
excitations can be gapped away from half filling.

Compared with the work of Choy and Phillips,\cite{Phillips} the
slave-rotor theory seems to have some similarities. The effective
"charge" chemical potential $\mu - \varphi_{0}$, associated with
softening of local charge fluctuations, can lie in the gap away
from half filling in the large $u/t$ limit, similar to the
half-filled case, while the effective "spin" chemical potential
$\varphi_{0}$ shows the conventional behavior as the Fermi liquid
or the fermion dynamics in the U(1) slave-boson theory of the
$t-J$ model. Considering that the slave-rotor description can
capture local antiferromagnetic correlations of $J$, allowing
double occupancy seems to play a crucial role for the emergence of
the Mott insulator away from half filling.

\subsection{U(1) pair-rotor gauge theory of the Hubbard model}

Now we consider the easy-plane limit
$\vec{\Omega}_{i}\cdot\vec{\tau} \equiv \Phi_{i}^{R}\tau_{1}$,
$U_{i} \equiv \exp[i\phi_{1i}\tau_{1}]$, and ${\vec
a}_{ij}\cdot{\vec \tau} \equiv a_{1ij}\tau_{1}$. Then, we find
another effective U(1) gauge Lagrangian\cite{Canonical_method}
\bqa && L_{\eta} =
\sum_{i}\eta_{i}^{\dagger}(\partial_{\tau}\mathbf{I} -
i\Phi^{R}_{i}\tau_{1})\eta_{i} \nn && -
tF\sum_{\ij}(\eta_{i}^{\dagger}e^{i{a}_{1ij}\tau_{1}}\tau_{3}\eta_{j}
+ H.c.) , \nn && L_{\phi} =
\frac{1}{2u}\sum_{i}(\partial_{\tau}\phi_{1i} - \Phi_{i}^{R})^{2}
\nn && - 2tE\sum_{\ij}\cos(\phi_{1j}+\phi_{1i} - a_{1ij}) . \eqa
We note that Eq. (24) can be reduced to Eq. (6) with
$\vec{\Omega}_{i}\cdot\vec{\tau} \equiv \Phi_{i}^{R}\tau_{1}$
after the gauge transformation of $\Phi_{i}^{R} \rightarrow
\Phi_{i}^{R} + \partial_{\tau}\phi_{1i}$, $a_{1ij} \rightarrow
a_{1ij} + \phi_{1i} + \phi_{1j}$, and Eq. (7) without $\phi_{2i}$
and $\phi_{3i}$ are utilized. If the $\tau_{3}$ matrix is not used
in Eq. (14), the hopping term in $L_{\phi}$ vanishes, and Eq. (6)
cannot be recovered from Eq. (24). Ignoring gauge fluctuations,
and replacing $\Phi_{i}^{R}$ with $\Phi_{0}$ as the mean-field
approximation, we obtain the on-site pairing order parameter given
by $\Phi_{0} = iu\langle\eta_{i}^{\dagger} \tau_{1}\eta_{i}\rangle
+ \langle\partial_{\tau}\phi_{1i}\rangle$. It turns out to be zero
because double occupancy costs too much energy. Thus, there is no
phase transition in the fermion Lagrangian at half filling as the
case of the easy-axis limit.

To examine the boson Lagrangian in the mean-field level, we resort
to the large $N$ generalization as the U(1) slave-rotor
representation. Introducing the $N$-component rotor field
$Y_{i\sigma}$, we rewrite $L_{\phi}$ in Eq. (24) as \bqa && L_{Y}
=
\frac{1}{2u}\sum_{i\sigma}(\partial_{\tau}Y_{i\sigma}^{\dagger})(\partial_{\tau}Y_{i\sigma})
- tE\sum_{\ij\sigma}(Y_{i\sigma}^{\dagger}Y_{j\sigma}^{\dagger} +
Y_{j\sigma}Y_{i\sigma}) \nn && +
i\sum_{i}\lambda_{i}(\sum_{\sigma}|Y_{i\sigma}|^{2} - 1) , \eqa
where $\lambda_{i}$ is a Lagrange multiplier field for the
uni-modular constraint. If we represent Eq. (25) in terms of
$Y_{i\sigma} = R_{i\sigma} + iI_{i\sigma}$ ($Y_{i\sigma}^{\dagger}
= R_{i\sigma} - iI_{i\sigma}$), we obtain the following expression
for the mean-field Lagrangian of collective pair excitations \bqa
&& L_{Y} =
\frac{1}{2u}\sum_{i\sigma}[(\partial_{\tau}R_{i\sigma})^{2} +
(\partial_{\tau}I_{i\sigma})^{2}] \nn &&
-2tE\sum_{\ij\sigma}(R_{i\sigma}R_{j\sigma} +
I_{i\sigma}I_{j\sigma}) +
i\sum_{i}\lambda_{i}[\sum_{\sigma}(R_{i\sigma}^{2} +
I_{i\sigma}^{2}) - 1] . \nn && \eqa One can find that Eq. (26) is
exactly the same as Eq. (18) if we rewrite Eq. (18) in terms of
$X_{i\sigma} = R_{i\sigma} + iI_{i\sigma}$. Thus, the mean-field
analysis in the previous slave-rotor theory can be directly
applied to Eq. (25). This leads us to conclude that both the
$\phi_{3i}$ and $\phi_{1i}$ fields are simultaneously incoherent
in $(u/t)>(u/t)_{0}$, and coherent in $(u/t)<(u/t)_{0}$ at half
filling, where $(u/t)_{0}$ is the critical value for the Mott
transition obtained in Eq. (23). This originates from $\varphi_{0}
= 0$ at half filling and $\mu_{c} = 0$ at the Mott-Hubbard
critical point, making Eq. (18) exactly the same as $L_{\phi}$ in
Eq. (24). Fundamentally, the reason why both fields should be
coherent simultaneously is the presence of the SU(2) pseudospin
symmetry at half filling. The slave-rotor action should be
symmetric (invariant) under the transformation $\phi_{1i}
\longleftrightarrow \phi_{3i}$ at half filling in the mean-field
approximation.

\section{Mean-field analysis of the SU(2) slave-rotor theory}

So far, we discussed that only two phases are expected to appear
at half filling in the context of the SU(2) slave-rotor theory,
corresponding to the spin liquid Mott insulator with
$\langle{e}^{i\phi_{3i}}\rangle = 0$,
$\langle{e}^{i\phi_{1i}}\rangle = 0$ and the Fermi liquid metal
with $\langle{e}^{i\phi_{3i}}\rangle \not= 0$,
$\langle{e}^{i\phi_{1i}}\rangle \not= 0$, respectively. There is
no intermediate phase because the presence of the SU(2) pseudospin
symmetry makes both phase fields simultaneously coherent or
incoherent. Away from half filling an intermediate phase,
characterized by condensation of only one kind of bosons, is
expected to arise because the SU(2) pseudospin symmetry is broken
explicitly due to hole doping, reflected in the chemical potential
term.

For the saddle-point analysis we consider the following mean-field
Lagrangian from Eq. (15), where SU(2) gauge fluctuations are
ignored ($e^{i\vec{a}_{ij}\cdot\vec{\tau}} \rightarrow I$), and
the easy-axis anisotropy for the SU(2) pseudospin order parameter
is naturally chosen ($\vec{\Omega}_{i}\cdot\vec{\tau} \rightarrow
- i \varphi_{0}\tau_{3}$), \bqa && L = L_{\eta} + L_{U} +
4t\sum_{\ij} EF , \nn && L_{\eta} =
\sum_{i}\eta_{i}^{\dagger}(\partial_{\tau}\mathbf{I} -
\varphi_{0}\tau_{3})\eta_{i} -
tF\sum_{\ij}(\eta_{i}^{\dagger}\tau_{3}\eta_{j} + H.c.) , \nn &&
L_{U} =
\frac{1}{4u}\sum_{i}\mathbf{tr}(-iU_{i}\partial_{\tau}U_{i}^{\dagger}
- i\varphi_{0}\tau_{3} + i\mu{U}_{i}\tau_{3}U_{i}^{\dagger})^{2}
\nn && -
tE\sum_{\ij}\mathbf{tr}(U_{j}^{\dagger}\tau_{3}U_{i}\tau_{3} +
H.c.) . \eqa The first assumption to ignore the SU(2) gauge
fluctuations is the simplest mean-field ansatz. The second
assumption of the easy-axis anisotropy for the SU(2) pseudospin
order parameter can be justified from the self-consistent
mean-field condition ${\vec \Omega}_{i} =
iu\langle\eta_{i}^{\dagger}\vec{\tau}\eta_{i}\rangle -
\frac{1}{2}\langle\mathbf{tr}[\vec{\tau}(-iU_{i}\partial_{\tau}U_{i}^{\dagger}
+ i\mu{U}_{i}\tau_{3}U_{i}^{\dagger})]\rangle$. The first term in
this expression becomes zero for $\tau_{1}$ and $\tau_{2}$ because
double occupancy costs too much energy. The second term also
vanishes for $\tau_{1}$ and $\tau_{2}$ because this term contains
pseudospin-flipping terms in the representation $U_{i} = \left(
\begin{array}{cc} z_{i\uparrow}  & - z_{i\downarrow}^{\dagger} \\
z_{i\downarrow} & z_{i\uparrow}^{\dagger}
\end{array} \right)$ of the SU(2)
slave-rotor matrix, but the boson Lagrangian $L_{U}$ has no
pseudospin-flipping terms as will be seen below. Particulary, the
second mean-field ansatz has important physical implication for
the SU(2) slave-rotor theory that the SU(2) pseudospin symmetry is
broken away from half filling.

Representing Eq. (27) in terms of the $z_{i\sigma}$ field, we
obtain the following expression  \bqa && L = L_{\eta} + L_{U} +
2t\sum_{\ij} EF , \nn && L_{\eta} =
\sum_{i}\eta_{i}^{\dagger}(\partial_{\tau}\mathbf{I} -
\varphi_{0}\tau_{3})\eta_{i} -
tF\sum_{\ij}(\eta_{i}^{\dagger}\tau_{3}\eta_{j} + H.c.) , \nn &&
L_{z} =
\frac{1}{2u}\sum_{i\sigma}(\partial_{\tau}z_{i\sigma}^{\dagger}-[\mu
-\sigma\varphi_{0}]z_{i\sigma}^{\dagger})(\partial_{\tau}z_{i\sigma}
+ [\mu -\sigma\varphi_{0}]z_{i\sigma}) \nn && -
tE\sum_{\ij\sigma}( \sigma{z}_{i\sigma}^{\dagger}z_{j\sigma} +
H.c.) + \lambda\sum_{i}(\sum_{\sigma}|z_{i\sigma}|^{2} - 1) , \eqa
where $2E$ and $\tau_{3}$ are replaced with $E$ and $\sigma =
\pm$, respectively. Although the SU(2) slave-rotor mean-field
Lagrangian Eq. (28) seems to be similar to the U(1) charge-rotor
mean-field Lagrangian Eq. (17) with Eq. (18), there is a crucial
difference between them. The most important ingredient in Eq. (28)
is the pseudospin-dependent chemical potential $\sigma\varphi_{0}$
in the boson sector. We show that this results in the condensation
of only one kind of bosons, allowing an intermediate phase away
from half filling. Since the renormalized dispersion of the
$z_{i\sigma}$ bosons depends on the pseudospin component, the
lowest energy of the $z_{i\uparrow}$ bosons lies at momentum
$(0,0)$ while that of the $z_{i\downarrow}$ bosons appears at
momentum $(\pi, \pi)$. There are no pseudospin-flipping terms in
the boson Lagrangian, thus the pseudospin-flipping terms such as
$\langle{z}_{i\uparrow}^{\dagger}\partial_{\tau}z_{i\downarrow}\rangle$
and $\langle{z}_{i\uparrow}^{\dagger}z_{i\downarrow}\rangle$
appearing in the self-consistent condition for the SU(2)
pseudospin order parameter $\vec{\Omega}_{i}$ should be zero,
justifying our assumption of the easy-axis anisotropy for the
order parameter.

Integrating out the $\eta_{i}$ and $z_{i\sigma}$ fields, we obtain
the mean-field free energy \bqa && F_{MF} = -
\frac{1}{\beta}\sum_{k,\omega}\ln[(i\omega)^{2} - (\varphi_{0} -
F\epsilon_{k}^{\eta})^{2}] \nn && + \frac{1}{\beta}
\sum_{k,\nu}\sum_{\sigma}\ln[-\frac{(i\nu + \mu -
\sigma\varphi_{0} )^{2}}{2u} + \lambda + E\sigma\epsilon_{k}^{z}]
\nn && + \sum_{k}(DEF - \mu\delta - \lambda) . \eqa Since the
$z_{i\downarrow}$ bosons have their energy minima at momentum $Q =
(\pi, \pi)$, we shift the momentum $k$ of the $z_{i\downarrow}$
boson to $k+Q$. Then, the mean-field free energy is given by \bqa
&& F_{MF} = - \frac{1}{\beta}\sum_{k,\omega}\ln[(i\omega)^{2} -
(\varphi_{0} - F\epsilon_{k}^{\eta})^{2}] \nn && + \frac{1}{\beta}
\sum_{k,\nu}\sum_{\sigma}\ln[-\frac{(i\nu + \mu -
\sigma\varphi_{0} )^{2}}{2u} + \lambda + E\epsilon_{k}^{z}] \nn &&
+ \sum_{k}(DEF - \mu\delta - \lambda) , \eqa where the $\sigma$
symbol disappears in the boson dispersion. Compared with the U(1)
charge-rotor free energy Eq. (19), Eq. (30) allows the
pseudospin-dependent chemical potential $\sigma\varphi_{0}$ in the
boson free energy. The charge chemical potentials appear to be
$\mu - \varphi_{0}$ for the $z_{i\uparrow}$ field and $\mu +
\varphi_{0}$ for the $z_{i\downarrow}$ field with the spin
chemical potential $\varphi_{0}$ in the SU(2) slave-rotor theory
while the charge chemical potential is given by $\mu -
\varphi_{0}$ for the $e^{i\phi_{3i}}$ field with the spin chemical
potential $\varphi_{0}$ in the U(1) charge-rotor theory.

Minimizing the free energy with respect to $E$, $F$,
$\varphi_{0}$, $\lambda$, and $\mu$, we find the self-consistent
saddle-point equations for all mean-field parameters \bqa && DE =
\int_{-D}^{D}d\epsilon{D}(\epsilon)\frac{1}{\beta}\sum_{\omega}\frac{2(\varphi_{0}
- F\epsilon)\epsilon }{(i\omega)^{2} - (\varphi_{0} -
F\epsilon)^{2} } , \nn && DF = -
\int_{-D}^{D}d\epsilon{D}(\epsilon)\frac{1}{\beta}\sum_{\nu}\sum_{\sigma}\frac{\epsilon}{-\frac{(i\nu
+ \mu - \sigma\varphi_{0})^{2}}{2u} + \lambda + E\epsilon} , \nn
&&
\int_{-D}^{D}d\epsilon{D}(\epsilon)\frac{1}{\beta}\sum_{\omega}\frac{2(\varphi_{0}
- F\epsilon)}{(i\omega)^{2} - (\varphi_{0} - F\epsilon)^{2} } \nn
&& = -
\int_{-D}^{D}d\epsilon{D}(\epsilon)\frac{1}{\beta}\sum_{\nu}\sum_{\sigma}\frac{\sigma\frac{i\nu
+ \mu - \sigma\varphi_{0}}{u}}{-\frac{(i\nu +\mu -
\sigma\varphi_{0})^{2}}{2u} + \lambda + E\epsilon} , \nn && 1 =
\int_{-D}^{D}d\epsilon{D}(\epsilon)\frac{1}{\beta}\sum_{\nu}\sum_{\sigma}\frac{1}{-\frac{(i\nu
+ \mu - \sigma\varphi_{0})^{2}}{2u} + \lambda + E\epsilon} , \nn
&& - \delta =
\int_{-D}^{D}d\epsilon{D}(\epsilon)\frac{1}{\beta}\sum_{\nu}\sum_{\sigma}\frac{\frac{i\nu
+ \mu - \sigma\varphi_{0}}{u}}{-\frac{(i\nu + \mu -
\sigma\varphi_{0})^{2}}{2u} + \lambda + E\epsilon} . \eqa One can
see from the denominator of the boson propagator that condensation
of the $z_{i\sigma}$ bosons occurs when their excitation gap
closes, given by $- \mu_{c} + \varphi_{0c} = 0$ for the
$z_{i\uparrow}$ bosons at momentum $(0, 0)$ and $- \mu_{c} -
\varphi_{0c} = 0$ for the $z_{i\downarrow}$ bosons at momentum
$(\pi, \pi)$, respectively. For both bosons to be condensed
simultaneously, $\mu_{c} = 0$ and $\varphi_{0c} = 0$ should be
satisfied. Actually, this happens at half filling where both
bosons have exactly the same transition point given by
$\lambda_{c} = E_{c}D$ as a result of the SU(2) pseudospin
symmetry. Away from half filling the effective chemical potential
$\varphi_{0}$ becomes nonzero, causing the SU(2) pseudospin
symmetry breaking not spontaneously but explicitly because it is
pseudospin-dependent. As a result, the $z_{i\uparrow}$ bosons
should be condensed before the condensation of the
$z_{i\downarrow}$ bosons occurs. Only one kind of bosons can be
condensed in the intermediate range of $u/t$, corresponding to the
intermediate phase mentioned before.

Performing the Matsubara frequency summation and energy
integration in Eq. (31), we find the following expressions for the
mean-field parameters $E$, $F$, $\varphi_{0}$, $\lambda$, and
$\mu$ at zero temperature \bqa && E = \frac{1}{2}\Bigl[1 -
\Bigl(\frac{\varphi_{0}}{DF}\Bigr)^{2}\Bigr] , \nn && F =
\frac{1}{6(DE)^2} \Bigl[(2\lambda-DE)\sqrt{2u(\lambda+DE)} \nn &&
- (2\lambda-E\epsilon_{-})\sqrt{2u(\lambda+E\epsilon_{-})}\Bigr]
\nn && + \frac{1}{6(DE)^2}
\Bigl[(2\lambda-DE)\sqrt{2u(\lambda+DE)} \nn && -
(2\lambda-E\epsilon_{+})\sqrt{2u(\lambda+E\epsilon_{+})}\Bigr] ,
\nn && \varphi_{0}\Bigl(\mu + \frac{E}{F}u\Bigr) = 0 , \nn && 1 =
\frac{\sqrt{2u(\lambda + DE)} -
\sqrt{2u(\lambda+E\epsilon_{-})}}{2DE} \nn && +
\frac{\sqrt{2u(\lambda + DE)} -
\sqrt{2u(\lambda+E\epsilon_{+})}}{2DE} , \nn && \delta = 1 -
\frac{\lambda}{DE} +
\frac{1}{2DE}\Bigl[\frac{(\mu-\varphi_{0})^{2}}{2u} +
\frac{(\mu+\varphi_{0})^{2}}{2u} \Bigr] , \eqa where
$\epsilon_{\pm}$ is given by $\epsilon_{\pm} =
\frac{1}{E}\Bigl[\frac{(\mu\pm\varphi_{0})^{2}}{2u} -
\lambda\Bigr]$. One can see the validity of these equations
performing the limit of $\delta \rightarrow 0$. We obtain
$\varphi_{0} = 0$ at half filling from the third equation. This
causes $\mu = - \sqrt{2u(\lambda - DE)}$ in the fifth equation.
Using the expressions for $\epsilon_{\pm}$ with $\mu = -
\sqrt{2u(\lambda - DE)}$ and $\varphi_{0} = 0$, one can find
$\epsilon_{\pm} = - D$, reproducing the mean-field equations for
the half-filled case.\cite{Half_filling_Eqs}

As soon as holes are doped, the chemical potential $\mu$ jumps
from $\mu = - \sqrt{2u(\lambda - DE)}$ to $\mu = - (E/F)u$.
Decreasing the Hubbard interactions with this chemical potential,
one can find the Mott-Hubbard transition point, where the
$z_{i\uparrow}$ bosons begin to be condensed. The $z_{i\uparrow}$
condensation occurs when $\lambda_{c} + E_{c}\epsilon_{-} = 0$ is
satisfied. From the analytic expressions in Eq. (32) with
$\epsilon_{\pm}$, we obtain the following conditions \bqa && E_{c}
= \frac{1}{2}\Bigl[ 1 -
\frac{E_{c}^{2}}{F_{c}^{4}}\Bigl(\frac{u_{c}}{D}\Bigr)^{2}\Bigr] ,
\nn && 3F_{c}E_{c}^{2} = \Bigl(2\frac{\lambda_{c}}{D} -
E_{c}\Bigr)\sqrt{2\frac{u_{c}}{D}\Bigl(\frac{\lambda_{c}}{D}+E_{c}\Bigr)}
\nn &&
-\frac{u_{c}}{D}\frac{E_{c}}{F_{c}}\Bigl[3\frac{\lambda_{c}}{D}-2\frac{u_{c}}{D}
\Bigl(\frac{E_{c}}{F_{c}}\Bigr)^{2}\Bigr] , \nn && \Bigl(E_{c} +
\frac{u_{c}}{D}\frac{E_{c}}{F_{c}}\Bigr)^{2} =
2\frac{u_{c}}{D}\Bigl(\frac{\lambda_{c}}{D} + E_{c} \Bigr) , \nn
&& \frac{\lambda_{c}}{D} = (1-\delta)E_{c} +
\frac{u_{c}}{D}\Bigl(\frac{E_{c}}{F_{c}}\Bigr)^{2} , \eqa
determining the critical value $u_{c}/D$ for the Mott-Hubbard
transition in the SU(2) slave-rotor theory away from half filling.
We note that the above equations for the Mott critical point do
not recover the half-filled case owing to the chemical potential
jump. From these equations we find $u_{c}/D \approx 0.170$ at
$\delta = 0.010$, $u_{c}/D \approx 0.171$ at $\delta = 0.050$, and
$u_{c}/D \approx 0.173$ at $\delta = 0.100$. The value of
$u_{c}/D$ for the Mott critical point increases as hole
concentration becomes larger, consistent with our expectation.
Remember that this Mott critical point is defined by the
condensation of the $z_{i\uparrow}$ bosons while the
$z_{i\downarrow}$ bosons are gapped owing to $- \mu_{c} -
\varphi_{0c} > 0$. We identify this Mott critical point with
$(u/t)_{2\delta}$.

There is another critical point associated with the condensation
of the $z_{i\downarrow}$ bosons with
$\langle{z}_{i\uparrow}\rangle \not= 0 $. It is clear that both
the $z_{i\uparrow}$ and $z_{i\downarrow}$ bosons get condensed
when $\lambda_{c} + E_{c}\epsilon_{-} = 0$ ($-\mu_{c} +
\varphi_{0c} = 0$) and $\lambda_{c} + E_{c}\epsilon_{+} = 0$ ($-
\mu_{c} - \varphi_{0c} = 0$) are satisfied, i.e., $\mu_{c} = 0$
and $\varphi_{0c} = 0$. Then, we obtain $\lambda_{c} =
(1-\delta)DE_{c}$ from the last equation in Eq. (32). Inserting
this expression with $E_{c} = 1/2$ into the fourth equation in Eq.
(32), we find the second quantum critical point \bqa
\frac{{u}_{c}}{D} = \frac{1}{4(2-\delta)} . \eqa The actual values
are $u_{c}/D \approx 0.127$ at $\delta = 0.010$, $u_{c}/D \approx
0.128$ at $\delta = 0.050$, and $u_{c}/D \approx 0.132$ at $\delta
= 0.100$. At half filling Eq. (34) recovers the Mott critical
point $u_{c}/D = 0.125$ of the U(1) slave-rotor theory exactly. We
define this critical value as $(u/t)_{1\delta}$. Note that
$(u/t)_{1\delta} < (u/t)_{2\delta}$ is satisfied as expected.

In summary, there is an intermediate region $(u/t)_{1\delta} < u/t
< (u/t)_{2\delta}$ away from half filling where only the
$z_{i\uparrow}$ bosons are condensed, i.e.,
$\langle{z}_{i\uparrow}\rangle \not= 0$ and
$\langle{z}_{i\downarrow}\rangle = 0$. The phase in $u/t <
(u/t)_{1\delta}$ is characterized by
$\langle{z}_{i\uparrow}\rangle \not= 0$ and
$\langle{z}_{i\downarrow}\rangle \not= 0$, identified with the
Fermi liquid metal while the phase in $u/t > (u/t)_{2\delta}$ is
given by $\langle{z}_{i\uparrow}\rangle = 0$ and
$\langle{z}_{i\downarrow}\rangle = 0$, named as the spin liquid
Mott insulator. We emphasize again that the presence of the
intermediate phase in $(u/t)_{1\delta} < u/t < (u/t)_{2\delta}$
results from the pseudospin-dependent chemical potential
$\sigma\varphi_{0}$ due to hole doping, causing the pseudospin
SU(2) symmetry breaking. The mean-field phase diagram is
summarized in Fig. 1, where "FL", "NFL", and "SL" represent Fermi
liquid, non-Fermi liquid, and spin liquid, respectively. Notice
that the intermediate phase is identified with the non-Fermi
liquid metal in the phase diagram Fig. 1. In the following we
investigate the nature of this intermediate phase named as the
non-Fermi liquid metal.

\begin{figure}
\includegraphics[width=8cm]{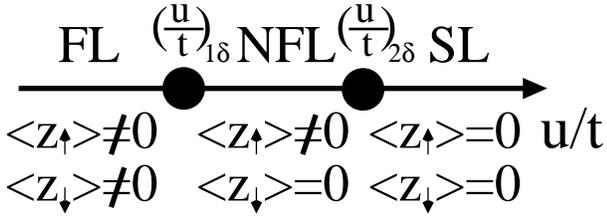}
\caption{\label{Fig. 1} A phase diagram in the SU(2) slave-rotor
theory of the Hubbard model away from half filling}
\end{figure}

One cautious person may suspect that the intermediate phase is an
artifact of gauge degrees of freedom. In other words, performing
gauge rotations to this ground state results in the same phase as
the Fermi liquid state. However, this guess is partially correct
only. It is important to discriminate the pseudospin SU(2)
symmetry from the SU(2) gauge symmetry. As discussed before, hole
doping causes the easy-axis anisotropy, breaking the SU(2)
pseudospin symmetry explicitly. However, the SU(2) gauge symmetry
cannot be broken.\cite{Elitzur} Unfortunately, the intermediate
phase breaks the SU(2) gauge symmetry. Is this phase a gauge
artifact? Not necessarily. This situation is exactly the same as
the characterization of superconductivity. Usually, we
characterize superconductivity as the condensed phase of Higgs
bosons (Cooper pairs). This identification necessarily breaks the
U(1) gauge symmetry, thus cannot be right in a rigorous manner. In
this respect there is no local order parameter to break the U(1)
gauge symmetry. However, we should admit that this
characterization is useful. The way to understand this
identification is to break the gauge symmetry explicitly, fixing
one gauge. Since the gauge symmetry is explicitly broken via gauge
fixing, one can define the local order parameter such as Cooper
pair bosons. Now we can understand the intermediate phase in the
same way. Consider gauge fixing first. Then, measure the
condensation of SU(2) rotor bosons. This is a physically appealing
way to discriminate various phases. In the following section we
show that the intermediate state is indeed different from the
Fermi liquid phase owing to the presence of incoherent fermion
pairs.

\section{Nature of the intermediate phase}

To reveal the nature of the intermediate phase, it is necessary to
find an effective Lagrangian. Low energy boson excitations are
considered to be \bqa && U_{i} = \left(
\begin{array}{cc} z_{i\uparrow}  & - z_{i\downarrow}^{\dagger} \\
z_{i\downarrow} & z_{i\uparrow}^{\dagger}
\end{array} \right) \approx \left(
\begin{array}{cc} \sqrt{x}  & - \sqrt{1-x}e^{-i\theta_{i}} \\
\sqrt{1-x}e^{i\theta_{i}} & \sqrt{x}
\end{array} \right) \nn \eqa for this ground state.
Here $x = |\langle{z}_{i\uparrow}\rangle|^{2}$ is the condensation
amplitude, not determined self-consistently in the previous
saddle-point analysis owing to complexity. $\theta_{i}$ is a
dynamic variable to guarantee partial freezing of charge
fluctuations, i.e., $\langle{z}_{i\downarrow}\rangle =
\sqrt{1-x}\langle{e}^{i\theta_{i}}\rangle = 0$. The ansatz of Eq.
(35) satisfies the uni-modular constraint
$\sum_{\sigma}|z_{i\sigma}|^{2} = 1$. Since the objective of this
section is to find a physical picture for this intermediate phase,
we do not try to determine the condensation amplitude $x$, and
just assume its presence.

The above ansatz for the SU(2) matrix field leads the
time-fluctuation term of the boson sector to be \bqa &&
\frac{1}{4u}\sum_{i}\mathbf{tr}(-iU_{i}\partial_{\tau}U_{i}^{\dagger}
+ \vec{\Omega}_{i}\cdot\vec{\tau} - i\varphi_{0}\tau_{3} +
i\mu{U}_{i}\tau_{3}U_{i}^{\dagger})^{2} \nn && =
\frac{1}{2u}\sum_{i} \Bigl[ (1-x)(\partial_{\tau}\theta_{i} +
\Omega_{zi} - i\varphi_{0} - i\mu)^{2} \nn && -
2\sqrt{x(1-x)}(\Omega_{xi}\cos\theta_{i} +
\Omega_{yi}\sin\theta_{i})(\partial_{\tau}\theta_{i} - 2i\mu) \nn
&& + 2(\Omega_{xi}^{2} + \Omega_{yi}^{2}) + 2x(\Omega_{zi} -
i\varphi_{0} + i\mu)^{2}  \Bigr] , \eqa where $\vec{\Omega}_{i}$
represents the time component of the SU(2) gauge field around its
zero mean-value. The $2x(\Omega_{zi} - i\varphi_{0} + i\mu)^{2}$
term allows us to replace $\Omega_{zi}$ with $i\varphi_{0} - i\mu$
in the low energy limit. This corresponds to the Anderson-Higgs
mechanism, as will be also seen in the boson hopping term.
Inserting $\Omega_{zi} = i\varphi_{0} - i\mu$ into Eq. (36), we
obtain the following expression for the right-hand-side (RHS) in
Eq. (36) \bqa && \mbox{RHS} \nn && =
\frac{1-x}{2u}\sum_{i}\Bigl(\partial_{\tau}\theta_{i} - 2i\mu -
\sqrt{\frac{x}{1-x}}[\Omega_{xi}\cos\theta_{i} +
\Omega_{yi}\sin\theta_{i}]\Bigr)^{2} \nn && + \frac{1}{u}\sum_{i}
\Bigl(\Omega_{xi}^{2} + \Omega_{yi}^{2} -
\frac{x}{2}[\Omega_{xi}\cos\theta_{i} +
\Omega_{yi}\sin\theta_{i}]^{2} \Bigr) . \nonumber \eqa One can see
$\Omega_{xi}$ and $\Omega_{yi}$ fluctuations gapped from \bqa &&
\frac{1}{u}\sum_{i} \Bigl(\Omega_{xi}^{2} + \Omega_{yi}^{2} -
\frac{x}{2}[\Omega_{xi}\cos\theta_{i} +
\Omega_{yi}\sin\theta_{i}]^{2} \Bigr) \nn && =
\frac{1}{u}\sum_{i}\Bigl( [1-x\cos^2\theta_{i}]\Omega_{xi}^{2} +
[1-x\sin^2\theta_{i}]\Omega_{yi}^{2} \nn && +
\frac{x}{2}[\Omega_{xi}\cos\theta_{i} -
\Omega_{yi}\sin\theta_{i}]^{2} \Bigr) , \nonumber \eqa thus safely
ignored in the low energy limit. As a result, we find the low
energy part for the time-fluctuation term \bqa &&
\frac{1}{4u}\sum_{i}\mathbf{tr}(-iU_{i}\partial_{\tau}U_{i}^{\dagger}
+ \vec{\Omega}_{i}\cdot\vec{\tau} - i\varphi_{0}\tau_{3} +
i\mu{U}_{i}\tau_{3}U_{i}^{\dagger})^{2} \nn && \approx \frac{
1-x}{2u}\sum_{i} (\partial_{\tau}\theta_{i} - 2i\mu)^{2} . \eqa

To obtain the low energy sector for the boson hopping term, we use
the following representation for the SU(2) gauge matrix $W_{ij}$
\bqa && W_{ij} = e^{-i\vec{a}_{ij}\cdot\vec{\tau}} \equiv \left(
\begin{array}{cc} X_{ij\uparrow}  & - X_{ij\downarrow}^{\dagger} \\
X_{ij\downarrow} & X_{ij\uparrow}^{\dagger}
\end{array} \right)  \eqa with the constraint $|X_{ij\uparrow}|^{2} + |X_{ij
\downarrow}|^{2} = 1$. Then, the boson hopping term is written as
\bqa && - t E \sum_{ij}
\mathbf{tr}(U_{j}^{\dagger}\tau_{3}W_{ij}U_{i}\tau_{3} ) \nn && =
- t E \sum_{ij}\Bigl[ x(X_{ij\uparrow} + X_{ij\uparrow}^{\dagger})
- (1-x)\{e^{-i\theta_{j}}X_{ij\uparrow}^{\dagger}e^{i\theta_{i}}
\nn && + e^{i\theta_{j}}X_{ij\uparrow}e^{-i\theta_{i}} \} -
\sqrt{x(1-x)} \{ X_{ij\downarrow}^{\dagger}e^{i\theta_{i}} +
X_{ij\downarrow}e^{-i\theta_{i}} \nn && +
X_{ij\downarrow}e^{-i\theta_{j}} +
X_{ij\downarrow}^{\dagger}e^{i\theta_{j}} \} \Bigr] . \eqa The
first term in the RHS makes the $X_{ij\uparrow}$ bosons condensed.
This is nothing but the Anderson-Higgs mechanism because the
condensation of the $z_{i\uparrow}$ bosons makes the gauge
fluctuations in the $X_{ij\uparrow}$ fields gapped. This
corresponds to the fact that the $\phi_{3i}$ condensation causes
gapped $a_{3ij}$ excitations in the U(1) slave-rotor gauge theory.
Considering the uni-modular constraint, we can set $X_{ij\uparrow}
= \sqrt{y}$ and $X_{ij\downarrow} = \sqrt{1-y}e^{ia_{ij}}$ as
$z_{i\uparrow} = \sqrt{x}$ and $z_{i\downarrow} =
\sqrt{1-x}e^{i\theta_{i}}$ of Eq. (35), where $y$ is the
condensation amplitude. Then, the boson hopping term is obtained
to be \bqa && - t E \sum_{ij}
\mathbf{tr}(U_{j}^{\dagger}\tau_{3}W_{ij}U_{i}\tau_{3} ) \nn && =
- t E \sum_{ij} \Bigl[ 2x\sqrt{y} - 2(1-x)\sqrt{y}\cos(\theta_{i}
- \theta_{j}) \nn && - 2\sqrt{x(1-x)(1-y)} \{ \cos(\theta_{i} -
a_{ij}) + \cos(\theta_{j} - a_{ij}) \} \Bigr] \nn && \rightarrow -
t E \sum_{ij} \Bigl[ 2x\sqrt{y} - 2(1-x)\sqrt{y}\cos(\theta_{i} -
\theta_{j}) \nn && - 2\sqrt{x(1-x)(1-y)} \{ \cos(\theta_{i} -
\theta_{j} - \tilde{a}_{ij}) + \cos \tilde{a}_{ij}  \} \Bigr] ,
\nn \eqa where we shift the gauge field $a_{ij}$ as
$\tilde{a}_{ij} = a_{ij} - \theta_{j}$ in the last line.

An important issue arises due to the $\cos \tilde{a}_{ij}$
potential term. One may claim that this term allows us to ignore
the gauge fluctuations at low energies because it causes
$\tilde{a}_{ij}$ excitations gapped. However, this problem does
not seem to be so simple owing to the stiffness parameter in the
$\cos$ term. When the condensation amplitudes $x$ and $y$ of the
$z_{i\uparrow}$ and $X_{ij\uparrow}$ bosons respectively are close
to $1$, the stiffness parameter of the $\cos$ potential becomes
very small. This may cause gauge fluctuations gapless. One can
argue that the $\cos$ potential would be always relevant in two
space and one time dimensions [$(2+1)D$] in the renormalization
group sense. However, this claim can be applied to the
conventional sine-Gordon model in $(2+1)D$.\cite{Kleinert} The
present problem is more complex since gauge fluctuations also
couple to fermion excitations, resulting in screening and
dissipation in the gauge dynamics.

First, we discuss the case when the $\cos \tilde{a}_{ij}$ term is
relevant. Then, we shift $\tilde{a}_{ij}$ as $\tilde{a}_{ij} +
\pi$ due to the minus sign of the $\cos$ potential, and set
$\tilde{a}_{ij} + \pi = a_{ij} - \theta_{j} + \pi = 0$ owing to
the relevance of the $\cos$ term. This is an important constraint
for gauge fluctuations. Because we consider the intermediate
mean-field phase characterized by $\langle{z}_{i\downarrow}\rangle
= \sqrt{1-x}\langle{e}^{i\theta_{i}}\rangle = 0$ and
$\langle{z}_{i\uparrow}\rangle = \sqrt{x}$, gauge fluctuations
cause $\langle{e}^{ia_{ij}}\rangle = 0$ that results from the
constraint $\langle e^{ia_{ij}}\rangle = \langle e^{i(\theta_{j} -
\pi)}\rangle$. As a result, the SU(2) gauge matrix $W_{ij}$ is
found to be $W_{ij} \approx \left(
\begin{array}{cc} \sqrt{y}  & \sqrt{1-y}e^{-ia_{ij}} \\ -
\sqrt{1-y}e^{ia_{ij}} & \sqrt{y}
\end{array} \right)$ with $\langle{e}^{ia_{ij}}\rangle = 0$. Note
that this SU(2) gauge matrix is consistent with the mean-field
analysis in Eq. (28) since this expression is reduced to the unit
matrix in the mean-field level owing to
$\langle{e}^{ia_{ij}}\rangle = 0$. We emphasize that the
off-diagonal components of the gauge matrix can arise in the
intermediate phase beyond the mean-field approximation, associated
with fermion pairing.

Using this gauge matrix, we find the following expression for the
fermion hopping term \bqa && - t F \sum_{ij}
\eta_{i}^{\dagger}W_{ij}^{\dagger}\tau_{3}\eta_{j} = - t F
\sum_{ij} \Bigl[ \sqrt{y}(\eta_{i+}^{\dagger}\eta_{j+} +
\eta_{j-}^{\dagger}\eta_{i-}) \nn && +
\sqrt{1-y}(\eta_{i+}^{\dagger}e^{-ia_{ij}}\eta_{j-}^{\dagger} +
\eta_{i-}e^{ia_{ij}}\eta_{j+}) \Bigr] . \eqa The key feature in
the fermion hopping term is that fermion pairing is dynamically
generated beyond the mean-field approximation due to the
off-diagonal components of the SU(2) gauge matrix although this
fermion pairing vanishes in the mean-field fashion owing to
$\langle{e}^{ia_{ij}}\rangle = 0$. This fermion hopping term
identifies the off-diagonal gauge fluctuations with phase
fluctuations of fermion pairs. Since the $\eta_{i}$ fermion
carries an electric charge owing to the condensation of the
$z_{i\uparrow}$ bosons, we interpret these dynamically generated
fermion pairs as preformed Cooper pairs owing to strong phase
fluctuations of the fermion pairs. An interesting point is that
the preformed pairs arise from the kinetic energy term instead of
the potential term in the Hubbard model.

The low energy effective Lagrangian for this phase is found to be
\bqa && L_{\eta} =
\sum_{i\sigma}\eta_{i\sigma}^{\dagger}(\partial_{\tau} -
\mu)\eta_{i\sigma} - t F \sqrt{y} \sum_{\ij\sigma}
(\eta_{i\sigma}^{\dagger}\eta_{j\sigma} + H.c.) \nn && - t F
\sqrt{1-y}\sum_{\ij}e^{-ia_{ij}}(\eta_{i+}^{\dagger}\eta_{j-}^{\dagger}
- \eta_{i-}^{\dagger}\eta_{j+}^{\dagger}) - H.c. , \nn &&
L_{\theta} = \frac{ 1-x}{2u}\sum_{i} (\partial_{\tau}\theta_{i} -
2i\mu)^{2} \nn && - 2t E \sum_{\ij} [\sqrt{x(1-x)(1-y)} -
(1-x)\sqrt{y}]\cos(\theta_{i} - \theta_{j}) , \nn \eqa where we
use $\Omega_{zi} = i\varphi_{0} - i\mu$ in the time-fluctuation
term for the $\eta_{i}$ fermions, obtained in the boson hopping
term. Remember the gauge constraint $\tilde{a}_{ij} + \pi = a_{ij}
- \theta_{j} + \pi = 0$. One can suspect this derivation because
the pairing symmetry is $s-wave$ instead of $d-wave$. In fact, the
pairing symmetry depends on the sign of the gauge matrix $W_{ij}$.
Thus, one can obtain the $d-wave$ fermion pairing from the kinetic
energy term considering $X_{ij\downarrow} =
-\sqrt{1-y}\varsigma_{ij}e^{ia_{ij}}$, where $\varsigma_{ij}$ is
$+$ when $j = i \pm \hat{x}$, and $-$ when $j = i \pm \hat{y}$. In
this case the low energy effective Lagrangian becomes \bqa &&
L_{\eta} = \sum_{i\sigma}\eta_{i\sigma}^{\dagger}(\partial_{\tau}
- \mu)\eta_{i\sigma} - t F \sqrt{y} \sum_{\ij\sigma}
(\eta_{i\sigma}^{\dagger}\eta_{j\sigma} + H.c.) \nn && - t F
\sqrt{1-y}\sum_{\ij}\varsigma_{ij}e^{-ia_{ij}}(\eta_{i+}^{\dagger}\eta_{j-}^{\dagger}
- \eta_{i-}^{\dagger}\eta_{j+}^{\dagger}) - H.c. , \nn &&
L_{\theta} = \frac{ 1-x}{2u}\sum_{i} (\partial_{\tau}\theta_{i} -
2i\mu)^{2} \nn && - 2t E \sum_{\ij}
[\varsigma_{ij}\sqrt{x(1-x)(1-y)} - (1-x)\sqrt{y}]\cos(\theta_{i}
- \theta_{j}) . \nn \eqa

It is not easy to determine which pairing symmetry will appear
without more sophisticated analysis. However, one can see that if
he compares the boson Lagrangian in Eq. (43) with that in Eq.
(42), the stiffness parameter in Eq. (43) is larger than that in
Eq. (42). In the case of $d-wave$ pairing we find $\rho_{x} = tE
|\sqrt{x(1-x)(1-y)} + (1-x)\sqrt{y}|$ and $\rho_{y} = tE
|\sqrt{x(1-x)(1-y)} - (1-x)\sqrt{y}|$, where $\rho_{x(y)}$ is the
stiffness parameter in the $x$ ($y$) direction. On the other hand,
we obtain $\rho_{x} = \rho_{y} = tE |\sqrt{x(1-x)(1-y)} -
(1-x)\sqrt{y}|$ in the case of $s-wave$ pairing. Because the
stiffness parameter in the $d-wave$ case is larger than that in
the $s-wave$ case, we expect that $d-wave$ pairing may be more
favorable than $s-wave$ pairing.

Note that if the condensation amplitude $x$ is close to $1$, the
assumption of $\langle{z}_{i\downarrow}\rangle =
\sqrt{1-x}\langle{e}^{i\theta_{i}}\rangle = 0$ is consistent with
the small stiffness parameter in Eq. (43). In this case the
resulting fermion Lagrangian becomes with the electromagnetic
vector potential $A_{ij}$ \bqa && L_{eff} =
\sum_{i\sigma}\eta_{i\sigma}^{\dagger}(\partial_{\tau} -
\mu)\eta_{i\sigma} \nn && - t F \sqrt{y} \sum_{\ij\sigma}
(\eta_{i\sigma}^{\dagger}e^{iA_{ij}}\eta_{j\sigma} + H.c.) \nn &&
- t F \sqrt{1-y}\sum_{\ij}\varsigma_{ij}e^{-ia_{ij}}
(\eta_{i+}^{\dagger}\eta_{j-}^{\dagger} -
\eta_{i-}^{\dagger}\eta_{j+}^{\dagger}) - H.c. \nn && -
\frac{1}{g^2}\sum_{\Box}\cos(\partial\times{a}) , \eqa where the
last gauge action is introduced to impose the condition
$\langle{e}^{-ia_{ij}}\rangle = 0$ with an internal gauge charge
$g$ of the $\eta_{i}$ fermion.

Eq. (44) is the main result of this section. Based on this
effective Lagrangian for the intermediate phase, we discuss its
physical implication. If quantum corrections due to gauge
fluctuations $a_{ij}$ are ignored as the mean-field approximation,
the effective fermion Lagrangian is reduced to \bqa && L_{eff} =
\sum_{i\sigma}\eta_{i\sigma}^{\dagger}(\partial_{\tau} -
\mu)\eta_{i\sigma} \nn && - t F \sqrt{y} \sum_{\ij\sigma}
(\eta_{i\sigma}^{\dagger}e^{iA_{ij}}\eta_{j\sigma} + H.c.) . \eqa
This seems to coincide with the Fermi liquid metal. Allowing the
density-rotor variable only as the U(1) slave-rotor
representation, the Fermi liquid metal appears when the
charge-rotor bosons are condensed.\cite{Florens} Actually, the
charge-rotor condensation makes the fermion field $\eta_{i}$
couple to the electromagnetic field $A_{ij}$. However, the
presence of the pair-rotor variable is expected to alter this
physical picture since the $\eta_{i\sigma}$ fermions are not
electrons but fractionalized ones due to gapped pair-rotor
excitations. To see the existence of coherent electron
excitations, we consider the electron spectral function given by
the convolution integral between the $\eta_{i\sigma}$ and
$z_{i\sigma}$ propagators   \bqa &&
G_{el\uparrow\uparrow}(ij,\tau\tau') =
\langle{T}_{\tau}[c_{\uparrow{i}\tau}c_{\uparrow{j}\tau'}^{\dagger}]\rangle
\nn && =
\langle{T}_{\tau}[U_{i\tau}^{\dagger}\eta_{i\tau}\eta_{j\tau'}^{\dagger}U_{j\tau'}]_{11}\rangle
\approx
x\langle{T}_{\tau}[\eta_{i\tau+}\eta_{j\tau'+}^{\dagger}]\rangle
\nn && +
(1-x)\langle{T}_{\tau}[e^{-i\theta_{i\tau}}e^{i\theta_{j\tau'}}]\rangle
\langle{T}_{\tau}[\eta_{i\tau-}^{\dagger}\eta_{j\tau'-}]\rangle .
\eqa There exist coherent electron excitations in the mean-field
fashion, resulting from the condensation of the $z_{i\uparrow}$
bosons. In this respect one can say that the intermediate phase
corresponds to the Fermi liquid state. But, it should be noted
that this result obtains in the saddle-point approximation
ignoring gauge fluctuations $a_{ij}$. The presence of gauge
fluctuations alters this picture completely. Notice how the Fermi
liquid renormalizes as a result of strong interactions. When $x =
1$, this renormalized Fermi liquid metal becomes identical with
the Fermi gas. Away from $x = 1$ we find the quasiparticle weight
$Z_{el} \sim x$, reduced by the presence of gapped pair-rotor
excitations, where the remaining portion $1-x$ is transferred into
incoherent backgrounds.

Strong gauge fluctuations corresponding to phase fluctuations of
fermion pairs should be allowed. Then, we expect that this phase
can be identified with a non-Fermi liquid metal owing to pairing
fluctuations. The presence of pairing-phase fluctuations is the
hall mark to discriminate this phase from the Fermi liquid metal,
represented as U(1) pair-gauge excitations. It is well known that
the presence of long-range gauge interactions can result in
non-Fermi liquid physics.\cite{Chubukov_NFL} When gauge
fluctuations are minimally coupled to gapless fermions, thus
screening of gauge fluctuations occurs via particle-hole
excitations, the effective U(1) gauge theory is characterized by
the dynamical critical exponent $z = 3$. In this case the
imaginary part of the fermion self-energy is given by
$\omega^{2/3}$ at the Fermi surface, implying that its real part
also has the same frequency dependence via the Kramer's Kronig
relation, thus giving rise to a non-Fermi liquid
behavior.\cite{Chubukov_NFL} The $\gamma$ coefficient of the
specific heat is proportional to $- lnT$ in three spatial
dimensions, and $T^{-1/3}$ in two dimensions.\cite{Kim_Kondo} The
dc conductivity is proportional to $T^{-5/3}$ in three dimensions
and $T^{-4/3}$ in two dimensions.\cite{Gauge_dynamics} However,
this non-Fermi liquid physics cannot be applied to the present
case because screening of gauge fluctuations arises in
particle-particle excitations instead of the particle-hole
channel. Unfortunately, we don't know the role of pair-gauge
fluctuations in the non-Fermi liquid physics at present. This
should be investigated near future.

On the other hand, if the $\theta_{i}$ fluctuations are
suppressed, i.e., $\langle{z}_{i\downarrow}\rangle =
\sqrt{1-x}\langle{e}^{i\theta_{i}}\rangle \not= 0$, the
corresponding gauge fluctuations would be also suppressed owing to
the gauge constraint. In this case the resulting effective
Lagrangian is obtained to be \bqa && L_{eff} =
\sum_{i\sigma}\eta_{i\sigma}^{\dagger}(\partial_{\tau} -
\mu)\eta_{i\sigma} \nn && - t F \sqrt{y} \sum_{\ij\sigma}
(\eta_{i\sigma}^{\dagger}e^{iA_{ij}}\eta_{j\sigma} + H.c.) \nn &&
- t F \sqrt{1-y}\sum_{\ij}\varsigma_{ij}
(\eta_{i+}^{\dagger}\eta_{j-}^{\dagger} -
\eta_{i-}^{\dagger}\eta_{j+}^{\dagger}) - H.c. \eqa This is
nothing but the effective Lagrangian for the $d-wave$ BCS
superconductivity. Thus, the $d-wave$ BCS superconductivity can
result from softening of pairing excitations in the SU(2)
slave-rotor gauge theory. The transition nature from the anomalous
metal with incoherent pairing to the $d-wave$ superconductor is
beyond the scope of this paper.

The SU(2) slave-rotor gauge theory can find the Fermi liquid,
superconductivity, non-Fermi liquid, and spin liquid in principle.
The Fermi liquid is described by $0 <
\langle{z}_{i\uparrow}\rangle = \sqrt{x} \leq 1$ with any
$\langle{z}_{i\downarrow}\rangle$ and
$\langle{X}_{ij\uparrow}\rangle = \sqrt{y} = 1$, $X_{ij\downarrow}
= 0$ ($W_{ij} = I$). The first condition results in the coherent
electron-quasiparticle weight given by Eq. (46). The second
condition indicates the absence of both diagonal and off-diagonal
fluctuations of the SU(2) gauge matrix fields, where the absence
of diagonal gauge fluctuations originates from the Anderson-Higgs
mechanism due to the $z_{i\uparrow}$ condensation while the
complete suppression of off-diagonal gauge fluctuations should be
determined by the self-consistent mean-field analysis, not
performed in this paper. In our mean-field analysis Eq. (28) both
the diagonal and off-diagonal bosons are condensed, i.e.
$\langle{z}_{i\sigma}\rangle \not= 0$, and the complete
suppression of the off-diagonal components in the SU(2) gauge
matrix is assumed, i.e., $W_{ij} = I$.

Superconductivity appears when $0 < \langle{z}_{i\uparrow}\rangle
= \sqrt{x} < 1$, $0 < \langle{z}_{i\downarrow}\rangle = \sqrt{1-x}
< 1$ and $0 \leq \langle{X}_{ij\uparrow}\rangle = \sqrt{y} < 1$,
$0 < \langle{X}_{ij\downarrow}\rangle = \sqrt{1-y} \leq 1$. The
presence of the off-diagonal $z_{i\downarrow}$ condensation is
necessary for the off-diagonal gauge bosons (${X}_{ij\downarrow}$)
to be condensed, causing coherent Cooper pairs for
superconductivity given by Eq. (47). Unfortunately, we could not
see superconductivity in our mean-field analysis because the
off-diagonal components in the SU(2) gauge matrix are not allowed
from the start. It is necessary to perform more sophisticated
mean-field analysis allowing the off-diagonal gauge components.

The anomalous metal arises when $0 < \langle{z}_{i\uparrow}\rangle
= \sqrt{x} < 1$, $\langle{z}_{i\downarrow}\rangle =
\sqrt{1-x}\langle{e}^{i\theta_{i}}\rangle = 0$ and $0 \leq
\langle{X}_{ij\uparrow}\rangle = \sqrt{y} < 1$,
$\langle{X}_{ij\downarrow}\rangle =
\sqrt{1-y}\langle{e}^{ia_{ij}}\rangle = 0$. One important thing is
that we cannot see the non-Fermi liquid phase in the mean-field
approximation ignoring gauge fluctuations. To obtain the anomalous
metallic phase, off-diagonal components should be allowed in the
SU(2) gauge matrix $W_{ij}$, meaning that the amplitudes of the
off-diagonal components should be nonzero, and the expectation
values of the off-diagonal components should be zero due to their
strong gauge fluctuations, not captured in the present
saddle-point analysis. What we have shown in our mean-field
analysis is that the intermediate phase characterized by partial
freezing of charge fluctuations, $\langle{z}_{i\uparrow}\rangle
\not= 0$ and $\langle{z}_{i\downarrow}\rangle = 0$, can appear
near the Mott-Hubbard critical point. We see that this
intermediate phase exhibits the Fermi liquid physics in the
saddle-point approximation. However, the pair-gauge fluctuations
(the off-diagonal gauge components) should be allowed, modifying
the Fermi liquid physics completely.
$\langle{z}_{i\downarrow}\rangle =
\sqrt{1-x}\langle{e}^{i\theta_{i}}\rangle = 0$ causes
$\langle{X}_{ij\downarrow}\rangle =
\sqrt{1-y}\langle{e}^{ia_{ij}}\rangle = 0$ self-consistently. As a
result, this intermediate phase is described by Eq. (44) with
incoherent pairing fluctuations.

The spin liquid Mott insulator is given by
$\langle{z}_{i\uparrow}\rangle = 0$,
$\langle{z}_{i\downarrow}\rangle = 0$ and
$\langle{X}_{ij\uparrow}\rangle = 0$,
$\langle{X}_{ij\downarrow}\rangle = 0$. This is consistent with
the mean-field analysis Eq. (28) except $X_{ij\downarrow} = 0$,
implying that this phase is identified with the U(1) spin liquid
Mott insulator in the context of our mean-field analysis. On the
other hand, if one admits fluctuations of the SU(2) gauge matrix
fields, this insulating phase is interpreted as the SU(2) spin
liquid Mott insulator due to SU(2) gauge fluctuations. In the case
of U(1) gauge fluctuations it was claimed that the U(1) spin
liquid Mott insulator can be stable against confinement resulting
from instanton excitations when there exist sufficiently large
flavors of gapless fermion
excitations.\cite{Hermele_ASL,Kim_SL_RG} However, in the case of
SU(2) gauge fluctuations the SU(2) spin liquid Mott insulator
would be unstable against confinement due to the SU(2) gauge
fluctuations\cite{Polyakov,Fradkin} although there are no reliable
calculations for this confinement problem owing to its complexity.
If the confinement is realized at low energies, the SU(2) spin
liquid is expected to turn into an antiferromagnetic insulator due
to particle-hole confinement. An interesting possibility can arise
that deconfined fermion excitations usually called spinons may
appear at high energies owing to the asymptotic
freedom\cite{Polyakov} of the SU(2) gauge theory. This may explain
the unidentified broad spin spectrum in the antiferromagnetic
phase of high T$_{c}$ cuprates, observed at high energies in the
inelastic neutron scattering measurements.\cite{Kim_DAF}

So far, we discussed the case when the $\cos \tilde{a}_{ij}$ term
is relevant. Now we consider the case when the $\cos
\tilde{a}_{ij}$ term is irrelevant. Shifting $\tilde{a}_{ij}$ as
$\tilde{a}_{ij} + \pi$ due to the minus sign of the $\cos$
potential, we obtain the SU(2) gauge matrix $W_{ij} \approx \left(
\begin{array}{cc} \sqrt{y}  & \sqrt{1-y}e^{-ia_{ij}} \\ -
\sqrt{1-y}e^{ia_{ij}} & \sqrt{y} \end{array} \right)$. An
important difference from the previous case is that there is no
gauge constraint for $a_{ij}$ and $\theta_{j}$ because the
$\cos\tilde{a}_{ij}$ term is assumed to be irrelevant. In this
case $\tilde{a}_{ij}$, $a_{ij}$, and $\theta_{j}$ are all strongly
fluctuating. The low energy effective Lagrangian is obtained to be
\bqa && L_{\eta} =
\sum_{i\sigma}\eta_{i\sigma}^{\dagger}(\partial_{\tau} -
\mu)\eta_{i\sigma} - t F \sqrt{y} \sum_{\ij\sigma}
(\eta_{i\sigma}^{\dagger}\eta_{j\sigma} + H.c.) \nn && - t F
\sqrt{1-y}\sum_{\ij}\varsigma_{ij}
e^{-ia_{ij}}(\eta_{i+}^{\dagger}\eta_{j-}^{\dagger} -
\eta_{i-}^{\dagger}\eta_{j+}^{\dagger}) - H.c. , \nn && L_{\theta}
= \frac{ 1-x}{2u}\sum_{i} (\partial_{\tau}\theta_{i} - 2i\mu)^{2}
\nn && - 2t E \sum_{ij}
[\varsigma_{ij}\sqrt{x(1-x)(1-y)}\cos(\theta_{i} - \theta_{j} -
\tilde{a}_{ij}) \nn && - (1-x)\sqrt{y}\cos(\theta_{i} -
\theta_{j})] . \eqa In the intermediate phase
$\langle{e}^{i\theta_{i}}\rangle = 0$ the fermion Lagrangian
$L_{\eta}$ in Eq. (48) recovers Eq. (44), thus the anomalous metal
is allowed even in this case. On the other hand, when the
$z_{i\downarrow}$ bosons become condensed, the $\tilde{a}_{ij}$
fluctuations would be suppressed due to the Anderson-Higgs
mechanism, thus $a_{ij}$ also, resulting in superconductivity
described by Eq. (47). However, this conclusion is in contrast
with the assumption that the $\cos\tilde{a}_{ij}$ term is
irrelevant. Thus, to be self-consistent between the assumption and
conclusion, the $\cos\tilde{a}_{ij}$ term should be relevant.

\section{Discussion and summary}

In this paper we try to answer to the question how to incorporate
Mott physics (charge fluctuations) in BCS-type superconductivity.
Particulary, the objective is to construct an effective theory
controlling pairing fluctuations in the BCS-like effective model.
Since the U(1) slave-rotor gauge theory allows local density
fluctuations only, this decomposition cannot control pairing
excitations. Remember that there are two kinds of gauge
fluctuations in the U(1) slave-rotor gauge theory with fermion
pairing [Eq. (3)], one of which corresponds to phase fluctuations
of fermion pairs. Although the U(1) charge-rotor variable governs
dynamics of density-gauge fluctuations, such boson excitations to
control pair-gauge fluctuations do not exist in the effective U(1)
gauge theory [Eq. (3)]. In this paper we find the missing
collective charge fluctuations represented as the pair-rotor
variable in the SU(2) slave-rotor gauge theory. The core of the
SU(2) slave-rotor gauge theory is that two kinds of collective
boson excitations are introduced to control two kinds of gauge
fluctuations. One of the boson excitations corresponds to the
density-rotor variable of the U(1) slave-rotor representation, and
the other is associated with on-site pairing fluctuations, called
the pair-rotor variable. One of the gauge fluctuations is
interpreted as the density-gauge field of the U(1) slave-rotor
representation, and the other is identified with phase degrees of
freedom for fermion pairs, called the pair-gauge field. Although
the SU(2) gauge structure is complex, we demonstrate that the
density-rotor variable controls the density-gauge field while the
pair-rotor variable governs the pair-gauge field. Since dynamics
of collective charge fluctuations can be controlled via the
local-interaction strength $u/t$, pairing fluctuations are
naturally handled.

We perform the saddle-point analysis based on the SU(2)
slave-rotor gauge theory. The crucial point in this analysis is
that the SU(2) pseudospin symmetry is explicitly broken via hole
doping, causing the easy-axis anisotropy. The easy-axis anisotropy
of the SU(2) pseudospin order parameter is reflected as the
pseudospin-dependent effective chemical potential for the SU(2)
slave-rotor variable, allowing that only local density
fluctuations are softened while local pairing fluctuations remain
gapped in the intermediate parameter range away from half filling.
Since density-rotor excitations are condensed, this phase exhibits
the Fermi liquid physics in the saddle-point approximation
corresponding to its nonzero quasiparticle weight proportional to
the condensation amplitude of the density-rotor bosons. However,
we see that this intermediate phase differs from the Fermi liquid
state beyond the mean-field approximation allowing gauge
fluctuations. Because pair-rotor excitations are gapped in this
phase, their corresponding pair-gauge excitations are strongly
fluctuating. Remember that the pair-gauge fluctuations can be
identified with phase fluctuations of fermion pairs. Thus,
incoherent fermion pairing does exist in this phase. The presence
of preformed fermion pairs discriminates this phase from the Fermi
liquid state completely [Eq. (44)].

It is valuable to discuss intuitively why local charge
fluctuations can be partially frozen away from half filling. At
half filling an on-site density fluctuation should induce an
on-site pair excitation. This is the origin why the density- and
pair-rotor bosons should be coherent or incoherent simultaneously.
Away from half filling the on-site density fluctuation need not
give rise to the on-site pair fluctuation because the density
fluctuations can occur between $|0\rangle$ and $|1\rangle$ sites,
but the on-site pair fluctuations should appear between
$|1\rangle$ and $|1\rangle$ sites, where $|0\rangle$ is an empty
site and $|1\rangle$ is one-electron site. As a result, away from
half filling the density-rotor bosons can be condensed against the
on-site Coulomb repulsion using the $|0\rangle$ and $|1\rangle$
sites while the pair-rotor bosons are gapped due to the on-site
Coulomb repulsion in the intermediate parameter range.

Several important issues remain open. The SU(2) slave-rotor gauge
theory opens the possibility of superconductivity in the Hubbard
model. Introducing the SU(2) slave-rotor matrix causes the SU(2)
gauge matrix naturally. Since the SU(2) gauge matrix allows its
off-diagonal components, off-diagonal fermion pairing can appear
naturally in the kinetic energy term. Unfortunately, we did not
perform a mean-field analysis to determine the amplitudes of the
off-diagonal components in the SU(2) gauge matrix owing to its
complex structure. More sophisticated self-consistent saddle-point
analysis should be performed near future for superconductivity in
the Hubbard model. Another problem is how to understand the spin
liquid Mott insulator away from half filling. Although the
slave-rotor gauge theory admits the presence of the Mott insulator
away from half filling owing to its special structure allowing
double occupancy, more extensive numerical simulations should be
performed for the Hubbard model. To understand the nature of the
non-Fermi liquid phase with preformed pairs such as transport and
thermodynamics, it is indispensable to find how to treat
pair-gauge fluctuations.

The stability of the non-Fermi liquid metal with incoherent
fermion pairing against disorder is an interesting open question
for this phase to be a genuine metallic phase in two dimensions.
The present author investigated the role of disorder in the two
dimensional fermion system with long-range gauge interactions,
where gauge fluctuations couple to charge currents instead of pair
currents as the present case.\cite{Kim_SR_disorder,Kim_exponent}
Long range interactions are shown to make the fermion system
stable against weak disorder even in two dimensions because the
gauge interactions let the fermions lie in a critical phase,
causing that the critical fermions feel their effective dimension
higher than two owing to their anomalous critical exponents.
Remember that there exists the localization-delocalization
transition in the weak disorder limit above two dimensions.
Criticality can protect fermions from localization due to disorder
in the weak disorder
limit.\cite{Kim_SR_disorder,Kim_exponent,NFL_exponent} However,
the present case should be addressed more thoroughly because gauge
fluctuations couple to pair currents instead of charge currents.
This problem may be related with the anomalous metallic behavior
in two dimensions.\cite{Review1,Review2}

K.-S. Kim would like to thank Dr. A. Tanaka for pointing out the
possibility for the SU(2) slave-rotor gauge theory to describe the
gossamer metal in Ref. \cite{Laughlin}. K.-S. Kim appreciates
helpful discussions with Drs. Y.-B. Kim, H.-Y. Kee, J.-W. Lee, K.
Park, J.-H. Han, and B. J. Yang. K.-S. Kim expresses thanks to Dr.
A. Rosch and his group members, particulary Dr. Markus for
illuminating discussions.


\begin{thebibliography}{9}
\bibitem{RMP} P. A. Lee, N. Nagaosa, and X.-G. Wen,
Rev. Mod. Phys. {\bf 78}, 17 (2006).
\bibitem{Trivedi} A. Paramekanti, M. Randeria, and N. Trivedi,
Phys. Rev. Lett. {\bf 87}, 217002 (2001); A. Paramekanti, M.
Randeria, and N. Trivedi, Phys. Rev. B {\bf 70}, 054504 (2004).
\bibitem{Laughlin} B. A. Bernevig, R. B. Laughlin, and D. I.
Santiago, Phys. Rev. Lett. {\bf 91}, 147003 (2003); Bogdan A.
Bernevig, George Chapline, Robert B. Laughlin, Zaira Nazario, and
David I. Santiago, cond-mat/0312573.
\bibitem{Florens} S. Florens and A. Georges, Phys. Rev.
B {\bf 70}, 035114 (2004).
\bibitem{LeeLee} S.-S. Lee and P. A. Lee, Phys. Rev. Lett. {\bf
95}, 036403 (2005).
\bibitem{Kim_SR_disorder} Ki-Seok Kim, Phys. Rev. B {\bf 73},
235115 (2006).
\bibitem{Slave_boson} The present decomposition scheme looks
somewhat similar to the slave-boson representation of the $t-J$
model. However, our decomposition scheme is nothing to do with the
slave-boson representation because the slave-boson representation
originates from how to solve the single occupancy constraint in
the $t-J$ model while the present decomposition does not. It
should be noted that this decomposition is not arbitrary since the
collective mode $U_{i\sigma\sigma'}$ is determined by its HS
field, as will be discussed in the text.
\bibitem{Kim_SU2_SRGT} Ki-Seok Kim, Phys. Rev. Lett. {\bf 97},
136402 (2006).
\bibitem{Kim_SR_doping} Ki-Seok Kim, Phys. Rev. B {\bf 74}, 115122
(2006).
\bibitem{Phillips} T.-P. Choy and P. Phillips, Phys. Rev. Lett.
{\bf 95}, 196405 (2005).
\bibitem{tJ_metal} K. Haule, A. Rosch, J. Kroha, and P. Wolfle,
Phys. Rev. B {\bf 68}, 155119 (2003); M. M. Zemljic and P.
Prelovsek, Phys. Rev. B {\bf 72}, 075108 (2005).
\bibitem{Half_filling_Eqs} At half filling ($\delta = 0$) we
obtain $\varphi_{0} = 0$ and $\mu = - \sqrt{2u(\lambda-DE)}$,
causing $\epsilon_{-} = -D$. As a result, we obtain the following
equations for the mean-field parameters $E$, $F$, and $\lambda$ at
half filling \bqa && E = \frac{1}{2} , ~~~~~ F =
\frac{1}{3(DE)^2}\Bigl[(2\lambda-DE)\sqrt{2u(\lambda+DE)} \nn && -
(2\lambda+DE)\sqrt{2u(\lambda-DE)}\Bigr] , \nn && 1 =
\frac{\sqrt{2u(\lambda + DE)} - \sqrt{2u(\lambda-DE)}}{DE} .
\nonumber \eqa Solving these algebraic equations, we find the
analytic expressions for the mean-field parameters \bqa && \lambda
= 2u + \frac{D^2}{32u} , ~~~~~~~ E = \frac{1}{2} , \nn && F =
\frac{(64u^2+D^2-8uD)(8u+D)-(64u^2+D^2+8uD)|8u-D|}{48D^2u} \nn &&
~~ = \frac{1}{24}\frac{D}{u} ~~~~~~~ \mbox{ for} ~~~ \frac{u}{D}
\geq \frac{1}{8} , \nn && ~~ =
\frac{64}{3}\Bigl(\frac{u}{D}\Bigr)^{2} ~~~~~~~ \mbox{for} ~~~
\frac{u}{D} < \frac{1}{8} . \nonumber \eqa Condensation of the
$\phi_{3i}$ bosons occurs when their excitation gap closes, given
by $\mu_{c} = 0$ or $\lambda_{c} - DE_{c} = 0$ that determines the
Mott-Hubbard critical point $u_{c}/D = 1/8$ in the mean-field
approximation.
\bibitem{Canonical_method} One can derive the pair-rotor theory
Eq. (24) in the canonical quantization method by decomposing the
electron Hilbert space into the composite Hilbert space of
$|\psi\rangle = |\eta\rangle\bigotimes|\Delta^{R}\rangle$ based on
the composite field representation $\psi_{i} =
e^{-i\phi_{1i}\tau_{1}}\eta_{i}$, where $\Delta^{R}$ represents
the density of an electron pair. This enlarged composite Hilbert
space is reduced to the original electron one by the constraint
$\Delta_{i}^{R} = \psi_{i}^{\dagger}\tau_{1}\psi_{i}$. This
canonical quantization can be realized in the following path
integral expression \bqa && Z = \int{D[\eta_{i}, \phi_{1i},
\Phi_{i}^{R}, \Delta_{i}^{R}]} \exp\Bigl[ -\int{d\tau}\Bigl\{
\sum_{i}\eta_{i}^{\dagger}\partial_{\tau}\eta_{i} \nn && -
t\sum_{ij}\eta_{i}^{\dagger}e^{i\phi_{1i}\tau_{1}}\tau_{3}e^{-i\phi_{1j}\tau_{1}}\eta_{j}
\nn && + \sum_{i}\Bigl(u\Delta_{i}^{R2} -
i\Delta_{i}^{R}\partial_{\tau}\phi_{1i} + i
\Phi_{i}^{R}(\Delta_{i}^{R} - \psi_{i}^{\dagger}\tau_{1}\psi_{i})
\Bigr)\Bigr\} \Bigr] , \nonumber \eqa resulting in the pair-rotor
theory Eq. (24) after integrating out the $\Delta_{i}^{R}$ field,
and performing the HS transformation for the hopping term. The
term $- i\Delta_{i}^{R}\partial_{\tau}\phi_{1i}$ leads us to
identify the $e^{-i\phi_{1i}\tau_{1}}$ operator with an
annihilation operator of an electron pair.
\bibitem{Elitzur} S. Elitzur, Phys. Rev. D {\bf 12}, 3978 (1975).
\bibitem{Kleinert} H. Kleinert, F. S. Nogueira, and A. Sudbo,
Phys. Rev. Lett. {\bf 88}, 232001 (2002); H. Kleinert, F. S.
Nogueira, and A. Sudbo, Nucl. Phys. B {\bf 666}, 361 (2003).
\bibitem{Chubukov_NFL} I. Vekhter and A. V. Chubukov, Phys. Rev.
Lett. {\bf 93} 016405 (2004).
\bibitem{Kim_Kondo} Ki-Seok Kim and Mun Dae Kim,
cond-mat/0608235, to be published in Phys. Rev. B.
\bibitem{Gauge_dynamics} P. A. Lee and N. Nagaosa, Phys. Rev.
B {\bf 46}, 5621 (1992).
\bibitem{Hermele_ASL} M. Hermele, T. Senthil, M. P. A. Fisher,
P. A. Lee, N. Nagaosa, and X.-G. Wen, Phys. Rev. B {\bf 70},
214437 (2004).
\bibitem{Kim_SL_RG} Ki-Seok Kim, Phys. Rev. B {\bf 72}, 245106
(2005).
\bibitem{Polyakov} A. M. Polyakov, \textit{Gauge Fields and Strings, Chap.
4} (Harwood Academic Publishers, 1987).
\bibitem{Fradkin} E. Fradkin, S. H. Shenker, Phys. Rev. D {\bf 19}, 3682
(1979).
\bibitem{Kim_DAF} {Ki-Seok Kim}, Phys. Rev. B {\bf 72},
214401 (2005).
\bibitem{Kim_exponent} Ki-Seok Kim, Phys. Rev. B {\bf 72}, 014406
(2005); Phys. Rev. B {\bf 70}, 140405(R) (2004).
\bibitem{NFL_exponent} S. Chakravarty, L. Yin, and E. Abrahams, Phys.
Rev. B {\bf 58}, R559 (1998).
\bibitem{Review1} E. Abrahams, S. V. Kravchenko, and M. P.
Sarachik, Rev. Mod. Phys. {\bf 73}, 251 (2001).
\bibitem{Review2} A. Kapitulnik, N. Mason, S. A. Kivelson, and S.
Chakravarty, Phys. Rev. B {\bf 63}, 125322 (2001).
\end{thebibliography}
\end{document}